\documentclass[letterpaper,journal]{IEEEtran}
\usepackage{amsmath,amssymb,amsfonts}
\usepackage{array}
\usepackage{textcomp}
\usepackage{stfloats}
\usepackage{url}
\usepackage{verbatim}
\usepackage{graphicx}
\def\BibTeX{{\rm B\kern-.05em{\sc i\kern-.025em b}\kern-.08em
		T\kern-.1667em\lower.7ex\hbox{E}\kern-.125emX}}
\usepackage{balance}
\usepackage{mdframed}
\usepackage{threeparttable}
\usepackage{multirow}
\usepackage{adjustbox}

\usepackage{pifont}
\usepackage{orcidlink}

\usepackage{algorithm}
\usepackage{algorithmicx}
\usepackage{algpseudocode}
\usepackage{microtype}
\usepackage{booktabs}

\usepackage{amsthm}
\usepackage{stmaryrd}

\usepackage[nocompress]{cite}

\usepackage{dsfont}
\usepackage{tabularx}
\usepackage{enumitem}
\newcolumntype{Y}{>{\centering\arraybackslash}X}
\usepackage[caption=false,font=scriptsize,
	labelfont=sf,textfont=sf]{subfig}
\usepackage[most]{tcolorbox}
\DeclareRobustCommand{\stage}[1]{%
	\ifmmode\text{\textsc{#1}}\else\textsc{#1}\fi%
}

\usepackage{xcolor}

\usepackage{colortbl}

\definecolor{bestgreen}{RGB}{169,208,142}
\definecolor{secondgreen}{RGB}{226,239,218}

\newcommand{\best}[1]{\cellcolor{bestgreen}\textbf{#1}}
\newcommand{\second}[1]{\cellcolor{secondgreen}#1}

\usepackage{xspace}
\makeatletter
\DeclareRobustCommand\onedot{\futurelet\@let@token\@onedot}
\def\@onedot{\ifx\@let@token.\else.\null\fi\xspace}
 
\def\ie{\emph{i.e}\onedot,~}

\makeatother

\usepackage{etoolbox}

\makeatletter
\renewcommand{\@IEEEBIOskipN}{2\baselineskip}

\expandafter\patchcmd\csname\string\IEEEbiography\endcsname
{\vskip \@IEEEBIOskipN plus 1fil minus 0\baselineskip}
{\vskip \@IEEEBIOskipN}
{}
{\PackageError{SecLog}{Could not patch IEEEbiography spacing}
	{Check IEEEtran version}}
\makeatother

\begin{document}
\title{SecLog-Bench: A Scalable Benchmark for Cybersecurity Log Reasoning with Scaffolded Evidence}

\author{Yifan Wang\orcidlink{0009-0008-4455-8264}, Dongliang Xu\orcidlink{0000-0002-6908-253X}, Xiaoyong Li\orcidlink{0009-0005-9977-8174}, Jing Li\orcidlink{0000-0002-7027-5574}, Peter S. Pan\orcidlink{0009-0007-5995-8212}, and Yue Yao\orcidlink{0000-0002-9852-4667}\\
\thanks{This work has been submitted to the IEEE for possible publication. Copyright may be transferred without notice, after which this version may no longer be accessible.}
\thanks{This work was supported by the Key Research and Development Program of Shandong Province, China (2025CXGC010901), the Shandong Province Overseas Young Talents Program (2026HWYQ-009), and the Shandong Key Laboratory of Industrial Network Security (2025CXPT069).}
\thanks{Y. Wang, D. Xu, X. Li, and Y. Yao are with the School of Airspace Science and Engineering, Shandong University, Shandong 264209, China. \textit{(Corresponding author: Dongliang Xu. E-mail: xudongliang@sdu.edu.cn.)}}

\thanks{J. Li is with the School of Computer Science, University of Technology Sydney, Ultimo, NSW 2007, Australia.}%
\thanks{P. S. Pan is an independent researcher in Braddon, ACT 2612, Australia.}%

}

\markboth{}%
{SecLog-Bench}

\maketitle

\begin{abstract}

Large language models (LLMs) have shown strong potential for cybersecurity, leading to growing interest in LLM-based automated cybersecurity log analysis. However, current datasets (1) are limited in scale; (2) rely mainly on single-source evidence; and (3) have only binary or multi-class attack labels, without dense annotations of the reasoning process.
To address these limitations, SecLog-Bench has three key features. First, we build an extensible simulation platform that covers 29 attack scenarios, supports repeated execution and the integration of new scenarios, and enables scalable generation of cybersecurity logs with attack-stage labels. Second, we use multi-source logs, including audit logs, application logs, and network traffic, to jointly analyze cyber attacks. Third, we annotate each sample with the essential context for cybersecurity log reasoning and a structured step-by-step evidence chain. We evaluate eight mainstream LLMs on this benchmark for post-detection attack-progress reasoning.
Results show that, when provided with scaffolded evidence, current LLMs exhibit modest capability in attack-progress reasoning, with the best model achieving only 44.44\% accuracy and 45.07\% Macro-F1. Their reasoning remains unreliable; \ie the highest Stage Reasoning F1 (SRF1) score is only 54.87\%. Moreover, even correct predictions may be accompanied by flawed reasoning, and reasoning quality is often misaligned with prediction accuracy.

\end{abstract}

\begin{IEEEkeywords}
	multi-source logs, large language models, security log reasoning.
\end{IEEEkeywords}

\section{Introduction}
\label{sec:introduction}

\IEEEPARstart{C}{ybersecurity} logs provide important evidence for threat analysis and incident response in security operations. These logs record host behavior, application semantics, and network interactions as an attack unfolds, thereby supporting security operations such as anomaly detection and attack confirmation~\cite{kent2006logmanagement,kent2006forensic,mitre_attack}. For example, in post-detection threat analysis, such log evidence is needed to infer attack progress and support response decisions~\cite{nelson2025incidentresponse}. Specifically, logs help establish relationships among evidence items to support subsequent judgments~\cite{mitre_attack,caltagirone2013diamond}. Thus, there is growing interest in using large language models (LLMs) to automate cybersecurity log analysis and infer attack progress.

However, current datasets and task definitions for cybersecurity log analysis have three limitations. First, many public datasets are released as fixed collections rather than reproducible data-generation environments~\cite{zhu2023loghub,sharafaldin2018toward,moustafa2016unsw,cse_cic_ids2018}. This fixed design limits controlled scenario variation, evidence regeneration, and expansion to emerging attack behaviors. Second, existing resources often provide either a single-source view or multiple source-specific views that remain fragmented. For example, system-log datasets~\cite{zhu2023loghub} focus on system events, whereas network-intrusion datasets~\cite{sharafaldin2018toward,moustafa2016unsw,cse_cic_ids2018} center on network traffic. Because attack activity can leave complementary traces across system, application, and network sources, assessing attack progress requires organizing and correlating observations across these sources rather than treating suspicious events in isolation~\cite{hossain2017sleuth,gao2018aiql,zeng2021watson,peng2025autolabel}. Multi-source resources broaden the available evidence, but generally do not organize observations from different sources into aligned multi-source evidence that supports joint analysis of attack progress~\cite{berady2022pwnjutsu,landauer2022aitldsv2,peng2025autolabel}.

Third, current task definitions generally frame cybersecurity log analysis as binary anomaly detection or multi-class attack classification~\cite{zhu2023loghub,sharafaldin2018toward,moustafa2016unsw,cse_cic_ids2018}. Even multi-source resources with attack labels generally lack dense annotations explaining how visible evidence supports the final judgment~\cite{landauer2022aitldsv2,peng2025autolabel}. This limitation is especially consequential for LLMs: a model may reach the correct label through incomplete evidence use, unsupported assumptions, or logically flawed reasoning~\cite{jiang2025mmecot,jiang2026m3cotbench}. Labels alone cannot establish reasoning reliability. Evaluating LLM-based cybersecurity log analysis therefore requires examining whether the model's reasoning is supported by the available log evidence rather than relying solely on final-label~accuracy.

To address these limitations, we present SecLog-Bench, a benchmark for cybersecurity log reasoning that evaluates LLMs on attack-stage prediction and reasoning reliability over multi-source log evidence. First, an extensible simulation platform covers 29 reproducible attack scenarios and supports repeated execution and scenario extension, enabling controlled evidence regeneration and coverage expansion. Second, host audit logs, application logs, and network traffic are collected during each execution and aligned using lightweight auxiliary indices recorded by runtime-specific adapters. The aligned logs are organized into scaffolded multi-source evidence and compressed into evidence windows for model evaluation. Third, each window is paired with an attack-stage label and a structured step-by-step reference reasoning chain, which is withheld from the evaluated model and used only after inference by the reasoning-aware evaluation protocol to assess model-generated reasoning. We formulate the resulting task as window-level attack-stage reasoning: given an attack-related multi-source log evidence window, a model predicts \stage{preparation}, \stage{execution}, or \stage{result} when the visible evidence sufficiently supports a concrete stage; otherwise, it outputs the evidence-insufficiency label \stage{stage\_unclosed}. The model must also output its own evidence-referenced reasoning chain.

We apply the reasoning-aware evaluation protocol to eight representative LLMs on a label-balanced, human-reviewed evaluation set of 216 scaffolded multi-source evidence windows. Results show that, when provided with scaffolded evidence, current LLMs exhibit useful but still limited capability in attack-stage prediction: the best model reaches 44.44\% accuracy and 45.07\% Macro-F1. Their reasoning, however, remains unreliable: the highest Stage Reasoning F1 (SRF1) score is only 54.87\%, correct predictions may still be accompanied by flawed reasoning, and reasoning quality is frequently misaligned with prediction accuracy. These findings show that final-label accuracy alone can overstate model capability, and that evaluating cybersecurity log reasoning requires reasoning-aware evaluation rather than detection-style scoring.

The main contributions of this paper are as follows:

\begin{itemize}
	\item
	We introduce SecLog-Bench, which formulates post-detection attack-progress analysis as window-level attack-stage reasoning over multi-source log evidence, requiring stage prediction and evidence-referenced reasoning.

	\item
	We develop an extensible, simulation-driven construction pipeline covering 29 reproducible attack scenarios, collecting host audit logs, application logs, and network traffic while recording lightweight auxiliary indices for cross-source evidence alignment.

	\item
	We annotate compressed multi-source evidence windows with attack-stage labels and structured step-by-step reference reasoning chains, and evaluate eight representative LLMs on attack-stage prediction, reasoning quality, reasoning reliability, and prediction--reasoning consistency.
\end{itemize}

\section{Motivating Example}
\label{background}
\begin{figure}[!t]
	\centering
	\includegraphics[width=\columnwidth]{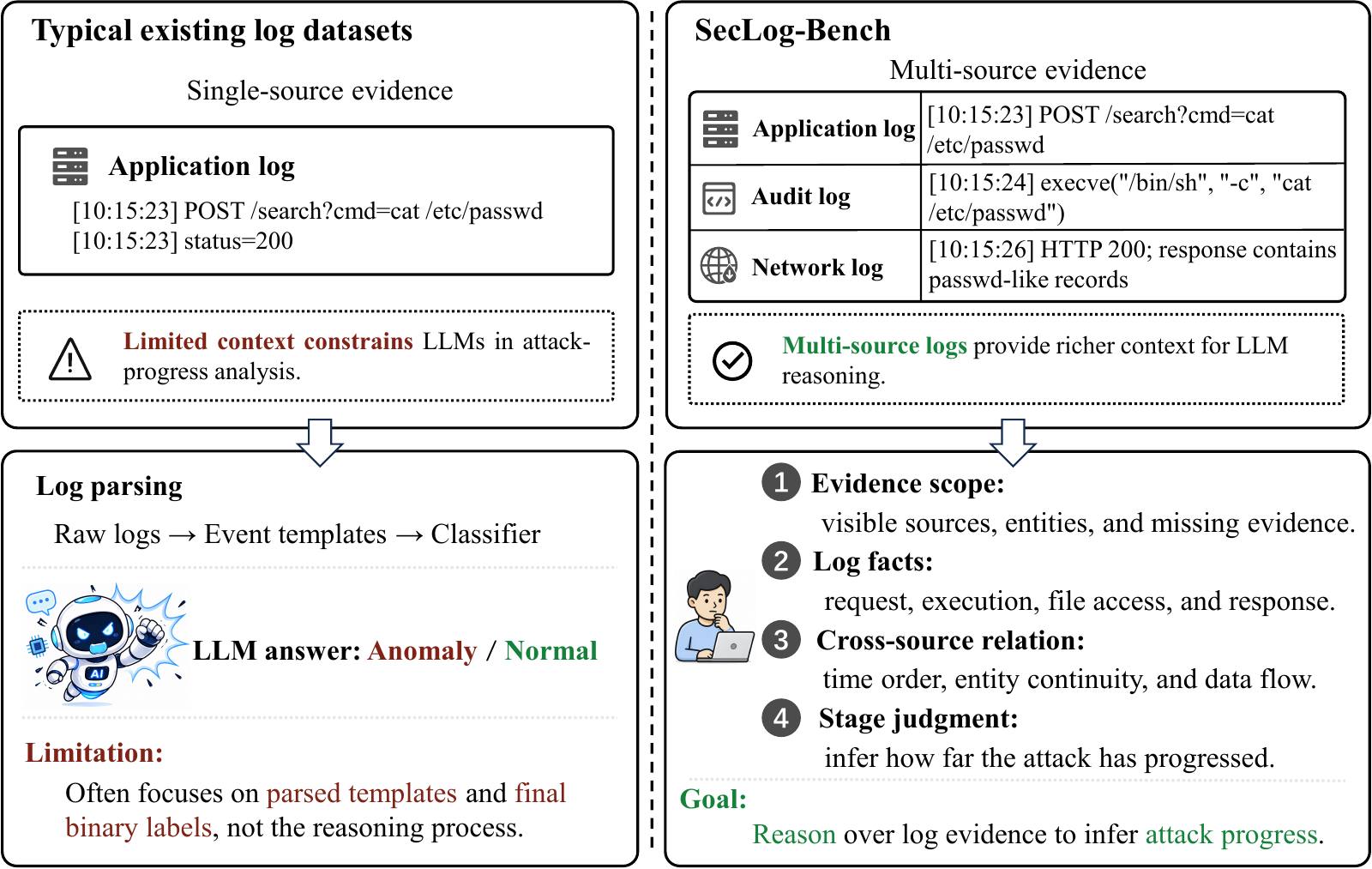}
		\vspace{-2ex}
	\caption{Motivation. Left: Typical log datasets provide only binary labels for anomalous and normal logs. Right: SecLog-Bench instead evaluates the reasoning process for attack-progress inference.
    }
	\label{fig:motivation}
		\vspace{-1ex}
\end{figure}

The following example illustrates how the evidence and annotation limitations discussed above affect attack-progress reasoning in practice. Fig.~\ref{fig:motivation} contrasts a conventional single-source view with the multi-source, reasoning-aware view used by SecLog-Bench.

On the left, the application log records a request containing \texttt{cat /etc/passwd} and an HTTP 200 status. Viewed alone, these records show that a suspicious request reached the application, but they do not establish command execution or returned output. The final ``Attack'' label therefore does not tell security analysts how far the attack progressed or which log records support that judgment~\cite{kent2006logmanagement,kent2006forensic,nelson2025incidentresponse}.

On the right, the application request is linked to an audit event showing command execution and a network response containing passwd-like records. Together, the three sources support a \stage{result} judgment; no single source establishes the full progression. SecLog-Bench organizes these observations into a traceable evidence window that preserves their temporal, entity, and data-flow relations. Each substantive claim in the reasoning chain must cite the visible evidence units on which it relies. This allows the benchmark to assess stage prediction, evidence support, and evidence-reference validity separately.

\section{Problem Definition}
\label{sec:problem-definition}

Motivated by the example above, we define the SecLog-Bench task as window-level attack-progress reasoning over scaffolded multi-source log evidence. In this benchmark, a model must infer the attack-stage label supported by the visible evidence and produce an evidence-referenced reasoning chain. Unlike conventional log anomaly detection, SecLog-Bench focuses on reasoning about attack progress within already identified attack-related evidence.

Because LLMs have finite context windows, we define the input to the log-analysis task as an evidence window $W$, which also serves as the evaluation unit. Rather than using a fixed-length time slice, a fixed-count log segment, or a sliding window, we select $W$ from the constructed attack-behavior chain to capture the evidence needed to characterize a stage-relevant portion of attack progress. A window is judged only on visible evidence within its boundaries; if that evidence supports no concrete stage, the model should output \stage{stage\_unclosed} rather than force a concrete label.

We define attack-stage labels based on the evidence visible within $W$ rather than directly adopting full kill-chain stages or ATT\&CK tactics~\cite{mitre_attack}. These taxonomies describe global attack progress, technical semantics, or behavior types, whereas SecLog-Bench evaluates whether a model can reason from multi-source evidence to an attack-stage judgment within $W$. Thus, a window label depends on the stage judgment supported by the current evidence, not on the global progress of the full attack scenario. We define $\mathcal{Y}=\{\stage{preparation},\allowbreak\stage{execution},\allowbreak\stage{result},\allowbreak\stage{stage\_unclosed}\}$ and abbreviate \stage{stage\_unclosed} as SU. The four labels are defined as follows:

\begin{itemize}[leftmargin=1em, labelsep=0.45em,
	itemsep=0pt, parsep=0pt, topsep=0pt, partopsep=0pt]
	\item \stage{preparation}: the window mainly shows request arrival, probing, or trigger conditions before the attack input is visibly handled by a server-side handler, runtime, or process.
	\item \stage{execution}: the window shows both the handling of the attack input by a server-side handler, runtime, or process and visible evidence of a corresponding attack-related operation.
	\item \stage{result}: the window contains observable consequences attributable to the attack behavior.
	\item SU: the window still belongs to the constructed attack chain, but the currently visible evidence is insufficient to reliably support any concrete stage label. SU is neither a benign class nor an actual stage in the attack lifecycle.
\end{itemize}
When cues for multiple stages are visible in the same window, the label follows the most advanced stage sufficiently supported by the visible evidence. Request arrival or probing alone does not support \stage{execution}. Visible server-side handling together with a corresponding attack-related operation may support \stage{execution}, whereas \stage{result} additionally requires observable consequences in the window.

Formally, each benchmark instance is represented as $I(W)=(E(W),X(W),y,R_{\mathrm{ref}}(W,y))$.
Here, $E(W)$ denotes the evidence associated with $W$, whereas the model-visible evidence view is constructed as
\[
X(W)=C_{\mathrm{model}}(E(W)),
\]
where $C_{\mathrm{model}}$ is a label-blind de-identification and compression procedure applied only to $E(W)$. The label $y\in\mathcal{Y}$ denotes the ground-truth stage label used to assess prediction correctness, while $R_{\mathrm{ref}}(W,y)$ denotes the human-confirmed reference reasoning chain used only after inference to assess reasoning quality. During inference, the model receives only $X(W)$, not $y$ or $R_{\mathrm{ref}}(W,y)$. The transformation $C_{\mathrm{model}}$ does not access $y$ or any information available only during annotation or review.

The model outputs $\hat{o}=(\hat{y},\hat{R})$, where $\hat{y}$ is the predicted attack-stage label and $\hat{R}=\{(\hat{\phi}_j,\hat{A}_j)\}_{j=1}^{m}$ is an evidence-referenced reasoning chain containing $m$ substantive claims. Here, $\hat{\phi}_j$ denotes the $j$-th substantive claim, and $\hat{A}_j$ denotes a nonempty set of explicit references cited in support of $\hat{\phi}_j$. Each reference identifies an indexed evidence unit in $X(W)$, facilitating evidence tracing by security analysts.

\begin{figure*}[!t]
	\centering
	\includegraphics[width=0.98\textwidth]{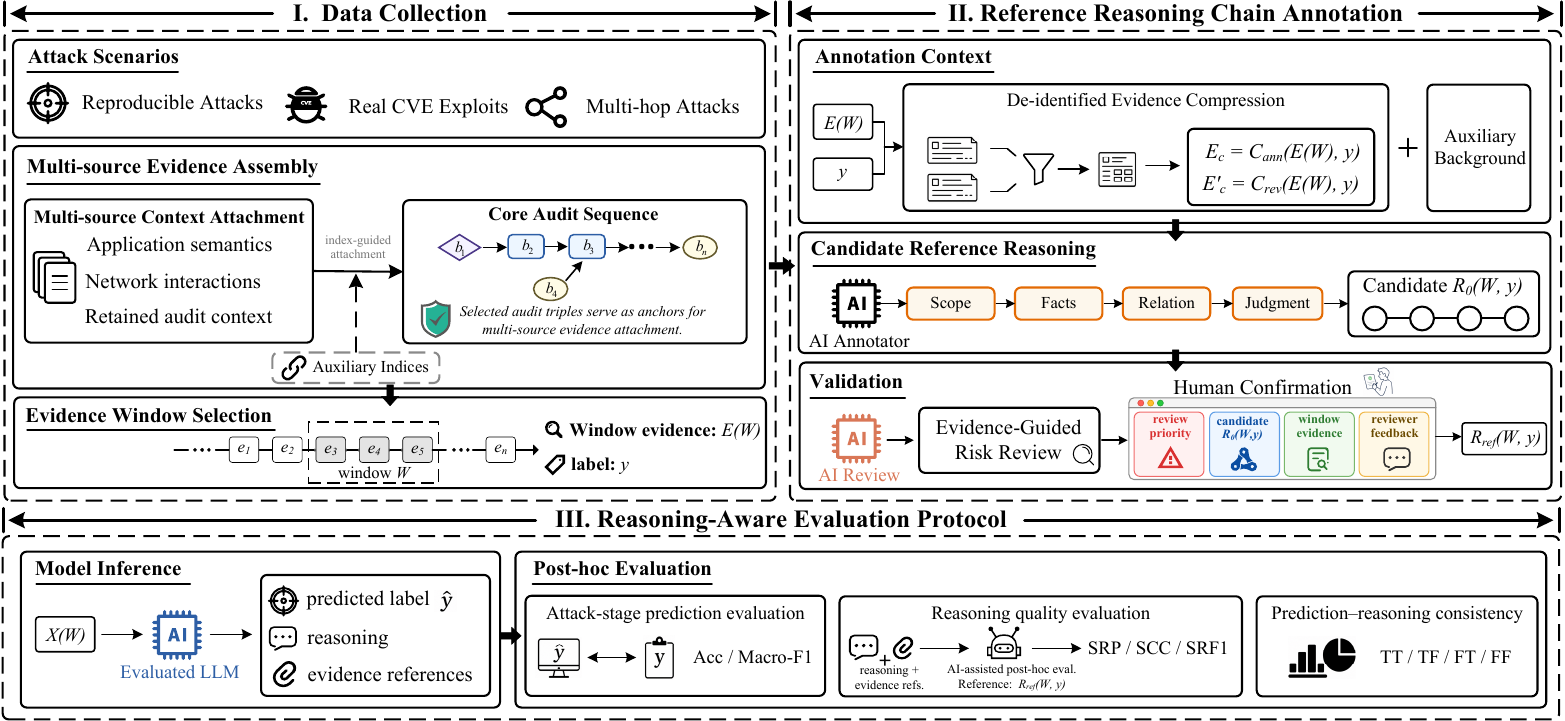}
	\vspace{-1ex}
	\caption{Overview of the SecLog-Bench workflow. The workflow constructs multi-source evidence windows, builds evaluation-only reference reasoning chains, and evaluates model attack-stage predictions and reasoning reliability.}
	\label{fig:overview}

\end{figure*}

\section{SecLog-Bench Construction}

This section presents the overall design of SecLog-Bench, which comprises three components: \emph{scalable multi-source attack-chain data collection}, \emph{reference reasoning chain annotation}, and a \emph{reasoning-aware evaluation protocol}.

\subsection{Workflow Overview}
\label{sec:overview-SecLog-Bench}
Fig.~\ref{fig:overview} presents the overall workflow of SecLog-Bench. Starting from reproducible attack scenarios, SecLog-Bench constructs multi-source log evidence windows for attack-stage reasoning. For each window, it retains the corresponding stage label and constructs a human-confirmed reference reasoning chain. During inference, the model receives only the model-visible evidence view $X(W)$. Post-hoc evaluation uses the stage label $y$ to assess attack-stage prediction and the human-confirmed reference reasoning chain $R_{\mathrm{ref}}(W,y)$ to assess reasoning reliability.

\subsection{Data Collection}
\label{sec:dataset-construction}

For each reproducible attack scenario, SecLog-Bench synchronously collects host audit logs, application logs, and network traffic. Using lightweight auxiliary indices recorded during execution, SecLog-Bench attaches application semantics, network interactions, and retained audit context to a core audit sequence constructed from host audit logs. The resulting attack-behavior chain is organized into stage-labeled source spans, from which local evidence windows are selected. Each selected window $W$ contains window evidence $E(W)$ and is associated with a stage label $y$.

\noindent\textbf{Attack Scenarios.}
We build the construction pool on a simulation platform covering 29 reproducible attack scenarios, including real CVE exploits, Sandworm-style simulations, and multi-hop attacks. These scenarios span multiple implementation stacks and application families, providing varied implementation contexts for evidence-window segmentation and attack-stage label derivation. We use Docker to create isolated network topologies and containers that emulate the corresponding hosts. The scenario hierarchy is presented in Section~\ref{sec:dataset-statistics}.

\noindent\textbf{Multi-source Evidence Assembly.}
To provide a common behavioral reference for organizing heterogeneous evidence across sources, SecLog-Bench adopts an audit-centered assembly process. For each specific attack scenario, we synchronously collect three distinct logs including host audit logs, application logs, and network traffic at runtime. Host audit logs record system-level behavior. Application logs record request paths, access status, handler execution, exception information, and response results to supplement server-side semantics. Network traffic records connection endpoints, communication direction, flows, and response-side exchanges to characterize network behaviors. Together, these three sources provide complementary evidence for attack-stage reasoning.

Host audit logs, in particular, provide system-level coverage of runtime behavior across processes, files, and network connections. They therefore serve as the source for constructing the core audit sequence. Auxiliary indices attach application semantics, network interactions, and useful non-core audit information as retained audit context to this sequence.

\noindent\textbf{Auxiliary Indices.}
Synchronous collection of these three sources alone is insufficient for reliable cross-source alignment. These sources have inconsistent time granularities, entity names, and observation perspectives. For example, a single application request may correspond to multiple system calls. Connection endpoints in network traffic may also fail to match host-side descriptions exactly because of container networking, NAT, or conntrack translation.

To bridge these gaps, we record a small set of auxiliary indices during scenario execution, including log-writer identity, request or processing-unit boundaries, traffic anchors, and NAT or connection aliases. This follows the idea of retaining runtime auxiliary information for multi-source log correlation in AutoLabel~\cite{peng2025autolabel}. However, our goal is not log-entry-level attack-relevance labeling, but window-level attack-stage reasoning. These indices neither change the attack behavior nor directly provide stage labels. Instead, they connect application logs and network traffic to the corresponding host behaviors represented in the core audit sequence, avoiding simple timestamp-based concatenation while preserving the traceability of raw evidence.

We implement this alignment mechanism with runtime-specific adapters for Python, PHP \& C, Java, and Node.js. The adapters target request handling and common system behavior rather than specific vulnerabilities or scenarios. Their context-capture and index-processing logic can therefore be reused across compatible scenarios that share the same runtime or language stack. Extending to such scenarios involves scenario-specific configuration and validation without redesigning the cross-source alignment workflow. For example, seven PHP scenarios share a request hook, while two GeoServer scenarios reuse the same Java agent and JNI artifacts. All 29 scenarios were integrated using these adapters.

\noindent\textbf{Core Audit Sequence.}
We construct the core audit sequence from host audit logs for later context attachment and window segmentation. Inspired by prior work on semantic representations of audit logs, such as WATSON~\cite{zeng2021watson} and UTLParser~\cite{tan2024unified}, we abstract low-level audit events into subject--relation--object audit triples that represent system-behavior relations among processes, files, and network entities. This representation abstracts low-level system calls into unified behavior units, reducing the impact of detailed audit fields, complex formats, and inconsistent semantics in raw audit logs. Our goal in using audit triples is to retain the stable behaviors needed to construct the core audit sequence, rather than to generate a general provenance graph or cluster all system events.

For audit normalization, we convert raw sysdig events into audit triples $t=(s,\rho,o)$, where the subject $s$ is a process, thread, or runtime entity, and the object $o$ is a file, child process, network connection, or remote endpoint. We group net, file, and process system calls into three relation families, NET, FILE, and PROCESS. The relation space is not constructed by simply retaining high-frequency syscalls. High-frequency events may indicate weakly causal operations and are not necessarily suitable for the relation set used to construct the core audit sequence. Therefore, we select relations by jointly considering event frequency, scenario coverage, category distribution, semantic stability, ability to capture causal or information-flow relations, and suitability for attack-stage reasoning. This design aggregates semantically similar syscall variants into stable relations while excluding weakly causal or high-noise events from the selected relation space.

We stratify audit triples according to their roles in constructing the attack chain. Audit triples representing stable behaviors are selected as core triples to form the core audit sequence; useful non-core audit information is preserved as retained audit context; and events with weak causality, high noise, or unstable categories are filtered or kept only as queryable records. The resulting traceable core audit sequence provides a stable basis for later multi-source context attachment without directly injecting all low-level events into model input.

\noindent\textbf{Multi-source Context Attachment.}
We assemble the multi-source evidence representation for attack-stage reasoning: application logs, network traffic, and retained audit context are attached to the core audit sequence using auxiliary indices for cross-source alignment and local structural relations. The goal is not to increase the number of log records, but to organize evidence scattered across different sources into the same behavioral context. As a result, each behavior represented in the core audit sequence retains its own system-level meaning while being linked to application-side semantics, network interactions, and retained audit context.

Multi-source evidence attachment follows structural cues rather than simple temporal proximity. For application logs, we use log-writing events, runtime units, and handler context to attach application-side events to the corresponding core audit edges. For network traffic, we follow the idea of traffic-to-audit-log correlation in AutoLabel and combine time, connection endpoints, NAT or conntrack aliases, and file-descriptor lifetimes to attach traffic events to the corresponding NET audit edges. Audit information that does not enter the core audit sequence but remains useful for explanation is preserved as retained audit context according to neighboring core audit edges, runtime units, entity matching, and relation types. This context helps explain the runtime setting in which the behaviors represented in the core audit sequence occur.

\begin{figure}[!t]
	\centering
	\includegraphics[width=\columnwidth]{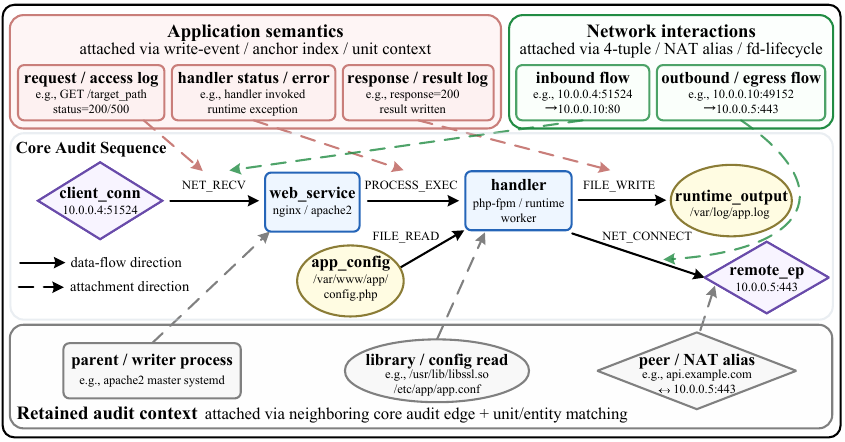}
	\vspace{-3.5ex}
    \caption{Example of multi-source context attachment within the window evidence $E(W)$. Application records, network flows, and retained audit context are attached to the core audit sequence through auxiliary indices and entity matching. Rounded rectangles, ovals, and diamonds denote processes, files, and sockets or network endpoints, respectively.}
	\vspace{-2.8ex}
	\label{fig:context-attachment-example}
\end{figure}

Fig.~\ref{fig:context-attachment-example} shows how application semantics, network interactions, and retained audit context are attached to the core audit sequence within an example evidence window. In this example, these auxiliary indices, together with local entity and context matches, link selected application- and network-side observations to concrete, traceable host behavior.

\noindent\textbf{Evidence Window Selection.} After multi-source context attachment, the resulting representation couples the core audit sequence with attached request semantics, server-side feedback, network interactions, and retained audit context. Evidence-source coordinates allow key evidence retained in the representation to be traced back to the corresponding raw log records. The attack-behavior chain is first organized into stage-labeled source spans along the core audit sequence, from which local evidence windows are selected. A selected window may retain the full source span or a stage-focused subinterval, with its boundaries chosen to preserve the stage-relevant behavioral sequence and its attached cross-source context needed for the associated stage judgment.

We denote a single evidence window as $W$, and define the visible and traceable core audit sequence, multi-source context, and corresponding evidence-source coordinates within $W$ as window evidence. The window evidence is denoted as $E(W)=\{e_i\}_{i=1}^{n}$, where $e_i$ denotes a traceable evidence unit within the window. $E(W)$ is used to verify the associated stage label and construct the reference reasoning chain.

For each evidence window, a preliminary stage label is inherited from its source span and independently verified or revised by two human reviewers against the criteria defined in Section~\ref{sec:problem-definition}. The final stage label $y$ must be supported by the visible and traceable evidence in $E(W)$ and the auditable behavioral relations within the current window boundary, rather than inferred from the global progress of the full attack scenario. Across all 216 evidence windows, the two reviewers achieved 91.67\% observed agreement on their stage-label decisions before adjudication, with Cohen's $\kappa=0.89$~\cite{cohen1960coefficient}; disagreements were resolved through consensus. In particular, a window is labeled SU when its visible attack-related evidence is insufficient to support a concrete stage judgment.

\subsection{Reference Reasoning Chain Annotation}
\label{sec:reference-reasoning-annotation}

Reference reasoning chain annotation constructs a human-confirmed $R_{\mathrm{ref}}(W,y)$ from window evidence $E(W)$ and stage label $y$.

A stage label $y$ only gives the final stage judgment for a window under the currently visible evidence. It does not explain which log facts and cross-source relations support the judgment, or why the current evidence is sufficient or insufficient for a concrete stage. The reference reasoning chain provides this missing explanation and serves as an evaluation reference for reasoning reliability assessment after model inference. The reference reasoning chain is not exposed to the evaluated model; after inference, it is used to determine whether reasoning units in the model output are supported by visible window evidence and cover key stage-judgment conditions.

\emph{First,} we construct a de-identified annotation context from the window evidence, retaining the evidence needed for attack-stage judgment while removing answer-revealing cues. De-identified auxiliary background helps the AI annotator interpret the relation between in-window behavior and attack mechanisms but cannot serve as evidence for the reasoning chain. \emph{Next,} with $y$ specifying only the annotation target, the AI annotator generates a candidate reference reasoning chain grounded in $E(W)$. \emph{Then,} an independent AI reviewer checks the candidate for evidence and stage-boundary risks, while an \emph{evidence-guided risk review} assigns higher review priority to high-risk candidates. \emph{Finally,} a candidate chain becomes the final reference reasoning chain only after \emph{human confirmation}.

\noindent\textbf{Annotation Context.}
Reference reasoning annotation takes the constructed window evidence $E(W)$ and its corresponding stage label $y$ as input. Here, $E(W)$ defines the window evidence scope within which the reference reasoning chain must be grounded. The label $y$ only constrains the annotation target, namely explaining why the window supports the given stage or why it can only yield an SU judgment. It cannot serve as reasoning evidence.

Within the annotation context, we apply \emph{de-identified evidence compression} as a rule-based process for constructing the annotation and review views from the window evidence $E(W)$. The process first removes information that may reveal the answer, including vulnerability identity, scenario identity, path cues, human-review traces, and existing reference reasoning content. It then selects, merges, and organizes behavioral facts, cross-source context, and source coordinates from $E(W)$ into these views. The process neither introduces facts outside the window nor rewrites the meaning of the original evidence.

Because attack-behavior windows often contain repeated system events, background activity, and evidence fragments from different sources, de-identified evidence compression must reduce context length while preserving the key evidence needed for stage judgment.
Compression therefore follows three selection principles, starting from the core audit triples selected during data collection in Section~\ref{sec:dataset-construction}:
\begin{itemize}[leftmargin=1.2em, labelsep=0.45em,
	itemsep=0pt, parsep=0pt, topsep=2pt, partopsep=0pt]
	\item \emph{Salience}: the process removes duplicate and weakly related events and prioritizes high-salience behaviors related to attack-chain progress, such as process execution, file access, network connections, and request handling.
	\item \emph{Cross-source continuity}: the process preserves cross-source relations that connect requests, processes, files, networks, and response results across host audit logs, application logs, and network traffic.
	\item \emph{Coverage}: evidence selection must cover the beginning and end of the window, key transition points, and the different sources while preserving stage-boundary cues. For SU windows, coverage additionally requires retaining the existing attack-related evidence and the visible reasons why that evidence is insufficient to support a concrete stage judgment, so that the window is not incorrectly represented as having no evidence.
\end{itemize}

Based on these compression principles, the annotation process constructs two label-conditioned views that are separate from the model-visible view $X(W)$: an annotation evidence view for candidate generation, $E_c=C_{\mathrm{ann}}(E(W),y)$, and a review evidence view, $E_c'=C_{\mathrm{rev}}(E(W),y)$. Here, $C_{\mathrm{ann}}(\cdot)$ and $C_{\mathrm{rev}}(\cdot)$ are both de-identified evidence compression functions, but they serve candidate generation and candidate review at different granularities. $E_c$ focuses on the key evidence needed to generate candidate reasoning, whereas $E_c'$ retains more window context and boundary cues for checking whether the candidate reasoning chain omits key evidence, contains over-inference, or has unclear attack-stage boundaries. Neither $E_c$ nor $E_c'$ is exposed to evaluated models. Importantly, all evidentiary content in both views must derive from $E(W)$, and neither view may introduce model predictions, out-of-window evidence, or the final reference reasoning chain.

The annotation evidence view can be further decomposed as $E_c=S_{\mathrm{ann}}(E(W),y)\cup G_{\mathrm{ann}}(E(W),y)$, where $S_{\mathrm{ann}}(E(W),y)\subseteq E(W)$ denotes raw evidence units selected from $E(W)$, and $G_{\mathrm{ann}}(E(W),y)$ denotes coverage or boundary descriptions deterministically generated from $E(W)$. The former ensures that candidate reasoning can return to concrete log facts. The latter presents source coverage, before-and-after relations of key behaviors, cross-source continuity, and visible evidence gaps that prevent a concrete stage judgment. $G_{\mathrm{ann}}(\cdot)$ only structures existing window evidence and does not constitute a new evidence source.

The annotation context additionally provides \emph{de-identified auxiliary background} to help the AI annotator understand the relation between in-window behavior and attack mechanisms. This design mirrors practical security analysis, in which analysts investigate logs with background knowledge of relevant vulnerabilities and affected components~\cite{kent2006logmanagement,kent2006forensic}. Auxiliary background may explain vulnerability mechanisms, affected-component semantics, or attack prerequisites, but it can only be used to interpret the behavioral semantics of $E(W)$ and cannot serve as evidence for the reasoning chain.

\noindent\textbf{Candidate Reference Reasoning.}
A reference reasoning chain must be constructed in an evidence-first manner: although $y$ constrains the annotation target, the chain never derives reasons from the label itself. Instead, reasoning starts from visible log facts within the window and gradually judges whether these facts can support the stage judgment. Candidate reasoning should return to the log facts themselves and also explain the behavioral relations among facts and their stage implications. These requirements call for a common organizational structure for candidate-chain generation, review, and human confirmation.

The four-step structure is abstracted from common assessment logic in security incident analysis rather than applied mechanically as a fixed template. Incident response and digital forensics typically require analysts to define the evidence scope and analysis boundary, confirm observable facts, correlate entities, behaviors, and data sources, and form evidence-supported analytical conclusions~\cite{kent2006logmanagement,kent2006forensic,nelson2025incidentresponse}. MITRE ATT\&CK~\cite{mitre_attack} organizes attack behavior from the perspectives of attack techniques and observable data sources, while the Diamond Model~\cite{caltagirone2013diamond} emphasizes entities and relations in intrusion activity. We therefore structure the reference reasoning chain as four steps: scope, facts, relation, and judgment.

Specifically, for a window $W$ and its stage label $y$, the reference reasoning chain is represented as:
\[
R_{\mathrm{ref}}(W,y)
=
(r_{\mathrm{scope}}, r_{\mathrm{facts}}, r_{\mathrm{relation}}, r_{\mathrm{judgment}}).
\]
Here, $r_{\mathrm{scope}}$, $r_{\mathrm{facts}}$, $r_{\mathrm{relation}}$, and $r_{\mathrm{judgment}}$ denote evidence-scope definition, observable-fact extraction, cross-source relation reasoning, and attack-stage judgment, respectively. Collectively, we use $r_j$ to denote any one of these four reasoning steps. For each $r_j$, let $A_j$ denote a nonempty set of explicit evidence references cited in support of that step, each identifying a traceable evidence unit in $E(W)$.

The four steps progress from evidence scope to attack-stage judgment. The step $r_{\mathrm{scope}}$ defines the evidence scope and analysis boundary by identifying visible sources, main entities, key behaviors, and missing or weak information, thereby constraining later reasoning to the current window evidence. The step $r_{\mathrm{facts}}$ extracts directly visible and verifiable log facts from the window. It preserves behavioral information across the three sources without prematurely reducing individual facts to a stage conclusion. The step $r_{\mathrm{relation}}$ places these facts back into the behavioral-chain context. It checks temporal adjacency, subject continuity, object consistency, and cross-source correspondence to determine whether scattered log facts form behavioral relations relevant to attack-stage judgment, while retaining weak links, evidence gaps, or boundary uncertainty. The step $r_{\mathrm{judgment}}$ forms the attack-stage judgment under these evidence constraints. It explains why the current visible evidence supports a concrete attack-stage judgment or, for SU, which attack-related cues are present and why those cues remain insufficient to support such a judgment.

To constrain the expression of reference reasoning chains, we require key reasoning statements to follow the sequence of facts in the evidence window $\rightarrow$ behavioral meaning $\rightarrow$ relation to the attack-stage judgment. Each statement first returns to visible log facts in the window, then explains their behavioral meaning in the attack-chain context, and finally states how they support a stage judgment or why they remain insufficient for a concrete stage judgment. This constraint prevents conclusions from being directly inferred from vulnerability background, scenario priors, or $y$, and avoids merely listing log facts without explaining their stage meaning.

The four-step reference reasoning chain is an evaluation reference rather than an output template: the evaluated model is not required to reproduce this structure or match the reference verbatim. As detailed in Section~\ref{sec:evaluation-methodology}, reasoning evaluation focuses on whether substantive reasoning units in the model output are supported by the current window evidence, cite valid evidence references, and cover key log facts, cross-source relations, and stage-judgment conditions.

SecLog-Bench uses an AI-assisted workflow for candidate annotation and risk review. AI assistance supports rather than replaces human annotation. Using the four-step structure above, the AI annotator receives the annotation evidence view $E_c$, de-identified auxiliary background, and $y$ to generate a candidate chain $R_0(W,y)$ grounded in $E(W)$. The independent AI reviewer identifies potential evidence risks and helps prioritize human review. In our implementation, GPT-5.5 and Claude Opus 4.7 serve as the AI annotator and independent AI reviewer, respectively, with access dates, prompts, and intermediate outputs recorded for auditability. The workflow is defined by candidate generation, independent risk checking, window-evidence constraints, and human confirmation rather than by the specific models used.

\noindent\textbf{Validation.}
During \emph{review}, the independent AI reviewer checks the candidate chain against the more complete review evidence view $E_c'$ for evidence omissions, unsupported inference, auxiliary-background substitution, stage-boundary errors, factual rewriting, and out-of-window dependence. The resulting feedback, risk flags, and suggested corrections are used in an \emph{evidence-guided risk review} to assign higher review priority to high-risk candidates. The AI reviewer does not regenerate the stage label, independently revise the candidate chain, or make the final acceptance decision.

Finally, a candidate chain becomes the final reference reasoning chain only after \emph{human confirmation}. Every candidate chain is manually checked through a dedicated review interface. Reviewers can jointly inspect the candidate chain, window evidence context, AI reviewer feedback, flagged risk items, and the resulting review priority. Samples with unclear boundaries, insufficient evidence support, or high review risk are further cross-checked by three reviewers. Human confirmation focuses on support from the current window evidence, the absence of dependence on auxiliary background or out-of-window information, and consistency with the stage criteria defined in Section~\ref{sec:problem-definition}. Across all 216 evidence windows, the pre-confirmation candidate-chain decisions produced by the AI-assisted annotation workflow show 92.13\% observed agreement with human consensus judgments, with Cohen's $\kappa=0.84$.

\subsection{Reasoning-Aware Evaluation Protocol}
\label{sec:evaluation-methodology}

The reasoning-aware evaluation protocol separates model inference from post-hoc evaluation, as illustrated in Fig.~\ref{fig:overview} (Part~III). During inference, the evaluated model receives only the model-visible evidence view $X(W)$ derived from $E(W)$, not $y$ or $R_{\mathrm{ref}}(W,y)$. Post-hoc evaluation then compares $\hat{y}$ with $y$, assesses reasoning against $R_{\mathrm{ref}}(W,y)$, and examines prediction--reasoning consistency.

\noindent\textbf{Model Inference.}
All models are evaluated on the same scaffolded compressed multi-source evidence view $X(W)$ defined in Section~\ref{sec:problem-definition}. Each model outputs a JSON object with a predicted label $\hat{y}$ and a natural-language reasoning chain whose substantive claims cite explicit evidence references.

\begin{figure}[!t]
	\centering
	\includegraphics[width=\columnwidth]{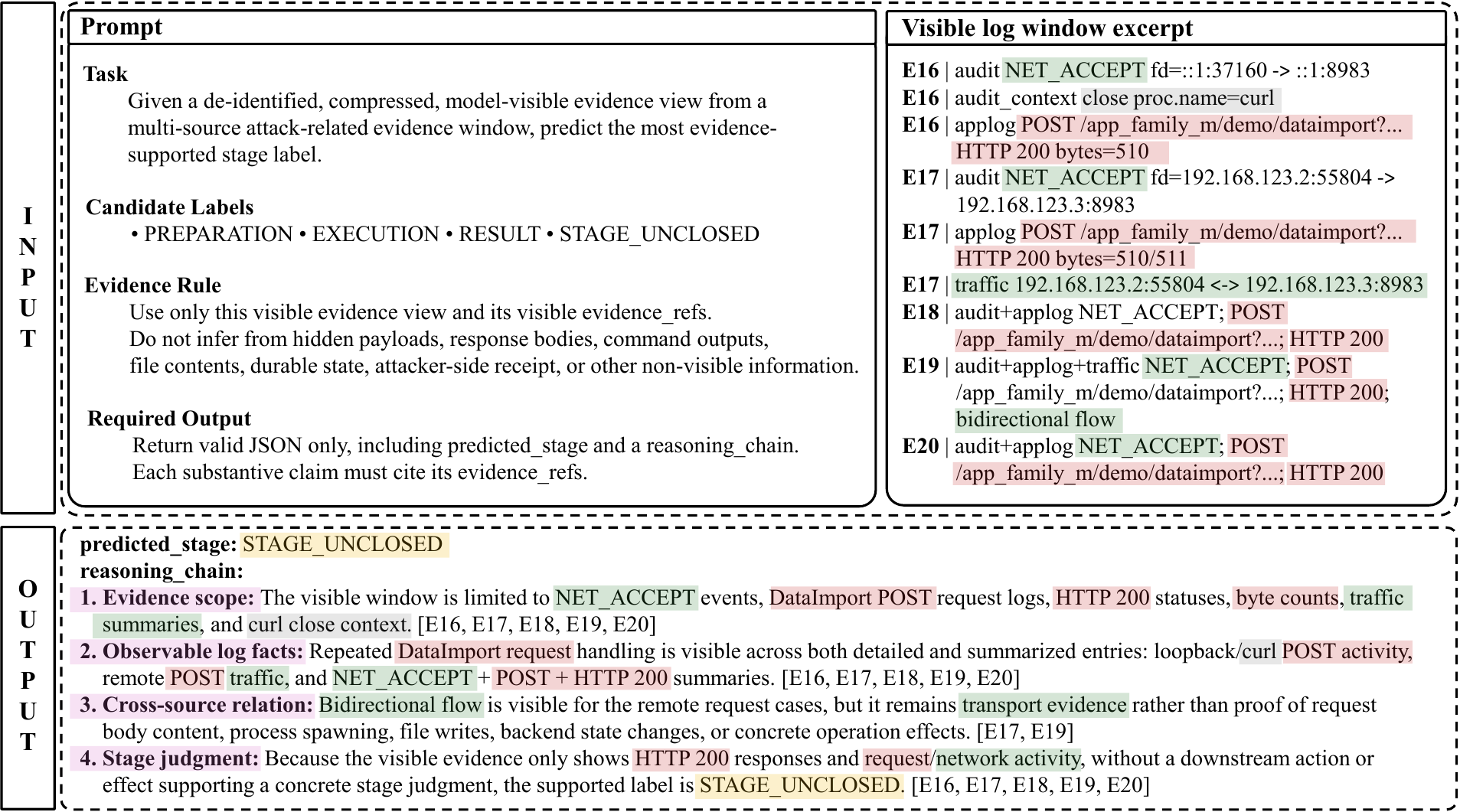}
	\vspace{-3ex}
	\caption{Example of model evaluation input and output in SecLog-Bench. Evidence highlights denote application semantics (pink), network interactions (green), and retained audit context (gray); yellow marks the predicted label. The example is shortened for readability.}
	\label{fig:evaluation-io-example}
	\vspace{-2ex}
\end{figure}

Fig.~\ref{fig:evaluation-io-example} illustrates this sample-level input-output format and the Evidence Rule, which restricts reasoning to $X(W)$ and requires each substantive claim to cite explicit references to indexed evidence units in $X(W)$. The four-step reasoning chain shown in the figure is illustrative; models may organize their reasoning chains freely within the required JSON output.

\noindent\textbf{Attack-Stage Prediction Evaluation.}
For attack-stage prediction, we compare the model prediction $\hat{y}$ with the ground-truth stage label $y$ and report Accuracy, Macro-F1, and per-label recall. Accuracy measures overall correctness, Macro-F1 reflects balance across the four labels, and per-label recall exposes label-specific weaknesses. If a model produces no parsable stage label or outputs an invalid one, the sample is counted as an incorrect prediction. Evidence-reference errors do not affect attack-stage prediction metrics; they are evaluated in the subsequent reasoning reliability assessment.

\noindent\textbf{Reasoning Quality Evaluation.}
We assess reasoning reliability post hoc with the identity of the evaluated model masked. The evaluation process first uses conservative local rules to decompose each model's reasoning output into separately verifiable reasoning units. Each unit is a substantive claim about log facts, behavioral relations, stage conditions, or reasons for insufficient evidence.

Building on prior reasoning-quality benchmarks that use independent LLM-based evaluators~\cite{jiang2025mmecot,jiang2026m3cotbench}, we use GPT-5.4-mini and Gemini 2.5 Pro as two independent evaluation assistants, neither of which is among the eight evaluated models. They perform structured checks of reasoning units using the same model-visible evidence view, reference reasoning chain, evaluation criteria, and output format. The checked dimensions include evidence support, evidence-reference validity, contradictions with visible evidence, reliance on out-of-window evidence, and over-inference. Evidence support measures whether a reasoning unit is directly supported by or reasonably inferred from visible evidence in the current window. Evidence-reference validity measures whether the model's evidence references come from the visible evidence scope of the current sample and are relevant to the corresponding reasoning unit.

Both assistants independently score every reasoning unit and key stage condition. For scalar items, the final value of each item is the mean of the two scores. If the two scores of an item differ by more than $0.30$, the item is flagged for high disagreement and manually reviewed, and the reviewed value is used in subsequent scoring. For binary risk items defined below, such as Invalid Evidence Reference and Critical Error, we apply a conservative rule: if either assistant triggers the item, it is retained in the reliability decision.

The assistants do not directly determine sample-level reliability; their outputs serve only as intermediate inputs, while fixed local scoring and aggregation rules, together with reliability thresholds, determine the final scores and reliability decisions.

We use Stage Reasoning Precision (SRP), Stage Condition Coverage (SCC), and Stage Reasoning F1 (SRF1) as reasoning quality metrics. SRP measures the precision of generated reasoning units under evidence constraints. For a single sample, let $\mathcal{L}=\{\ell_i\}_{i=1}^{N}$ denote the set of reasoning units extracted from the model's reasoning output, where $\ell_i$ is the $i$-th reasoning unit. When $N>0$, SRP is defined as
\begin{equation}
	\label{eq:srp}
	\begin{aligned}
		s(\ell_i)
		&=
		g_i q_i^{\mathrm{sup}} q_i^{\mathrm{ref}}
		p_i^{\mathrm{over}}
		q_i^{\mathrm{stage}} p_i^{\mathrm{bound}}, \\
		\mathrm{SRP}
		&=
		\frac{1}{N}\sum_{i=1}^{N}s(\ell_i).
	\end{aligned}
\end{equation}
If no verifiable reasoning unit can be extracted, we set $\mathrm{SRP}=0$.
Fixed local rules map the aggregated reasoning-unit judgments to the factors in $s(\ell_i)$, whereas key-stage-condition assessments are aggregated separately to compute SCC. The critical-error indicator $g_i$ is set to zero if the reasoning unit contains a severe evidence error, contradicts the current-window evidence, or depends on out-of-window evidence, and to one otherwise. The factors $q_i^{\mathrm{sup}}$, $q_i^{\mathrm{ref}}$, and $q_i^{\mathrm{stage}}$ score evidence support, evidence-reference validity, and support for the stage judgment, while $p_i^{\mathrm{over}}$ and $p_i^{\mathrm{bound}}$ penalize over-inference and boundary errors. Each of these five ordinal factors takes one of three fixed values in $[0,1]$ according to the corresponding aggregated judgment.

SCC measures whether the model's reasoning covers the key judgment conditions in the reference reasoning chain that are directly relevant to the sample's stage-label decision. For a sample, let $\mathcal{C}=\{c_j\}_{j=1}^{M}$ denote its set of $M$ key stage conditions, with $M>0$. We set $h(c_j)$ to $1.0$, $0.5$, or $0$ when the model's reasoning fully covers, partially covers, or does not correctly cover $c_j$, respectively. SCC is defined as
\begin{equation}
	\label{eq:scc}
	\mathrm{SCC}
	=
	\frac{1}{M}
	\sum_{j=1}^{M}
	h(c_j).
\end{equation}

SRF1 is an aggregate reasoning quality metric that combines the reasoning precision measured by SRP and the key stage-condition coverage measured by SCC. When $\mathrm{SRP}+\mathrm{SCC}=0$, we set $\mathrm{SRF1}=0$:
\begin{equation}
	\label{eq:srf1}
	\mathrm{SRF1}
	=
	\frac{2\cdot \mathrm{SRP}\cdot \mathrm{SCC}}
	{\mathrm{SRP}+\mathrm{SCC}}.
\end{equation}

We also report two sample-level diagnostic metrics: Critical Error and Invalid Evidence Reference. Critical Error indicates that at least one reasoning unit has $g_i=0$ due to a severe evidence error, a contradiction with current-window evidence, or dependence on out-of-window evidence. Invalid Evidence Reference indicates that a reference falls outside the current visible scope or does not match its reasoning unit.

At the model level, we average SRP and SCC across samples and compute SRF1 as their harmonic mean using Eq.~\ref{eq:srf1}. Separately, at the sample level, we judge reasoning as reliable if it satisfies $\mathrm{SRP}\geq 0.33$, $\mathrm{SCC}\geq 0.50$, and $\mathrm{SRF1}\geq 0.45$, and does not trigger Critical Error or Invalid Evidence Reference; otherwise, it is judged unreliable. The thresholds were calibrated on a pilot subset, disjoint from the validation subset below, to align with human reliability judgments and then fixed before the main evaluation. These thresholds define the Main policy, whose local sensitivity to the SRP threshold is examined in Section~\ref{sec:rq2-reasoning-quality}.

The ordinal mappings and multiplicative aggregation rule are fixed components of the evaluation rubric rather than learned parameters or reliability thresholds; they were specified before the main evaluation and applied uniformly to all model outputs. Human reviewers adjudicate individual cases with unclear stage boundaries or large disagreements between the two evaluation assistants, but they do not alter the rubric, aggregation rule, or reliability thresholds.

Finally, to validate the assistant-supported reliability assessment protocol, we conduct a human-only review of 44 evidence windows (about 20\% of the evaluation set, with 11 windows from each ground-truth label). The review covers 352 outputs from the eight evaluated models. Human evaluators judge sample-level reasoning reliability using the corresponding evidence window, reference reasoning chain, and reliability decision rules, without access to evaluation-assistant outputs or model identities. Human judgments show 89.20\% observed agreement with the sample-level decisions of the fixed evaluation pipeline, with Cohen's $\kappa=0.78$.

\noindent\textbf{Prediction--Reasoning Consistency.}
Based on the sample-level reasoning reliability decision above, we categorize each sample into TT, TF, FT, or FF to analyze the relation between attack-stage prediction correctness and reasoning reliability. The four categories denote correct prediction with reliable reasoning (TT), correct prediction with unreliable reasoning (TF), incorrect prediction with reliable reasoning (FT), and incorrect prediction with unreliable reasoning (FF), respectively. We additionally report $\mathrm{TT}/(\mathrm{TT}+\mathrm{TF})$ as TT/Correct and $\mathrm{FT}/(\mathrm{FT}+\mathrm{FF})$ as FT/Wrong.

\section{Benchmark Statistics}
\label{sec:dataset-statistics}

\begin{figure*}[!t]
	\centering
	\includegraphics[width=0.89\textwidth]{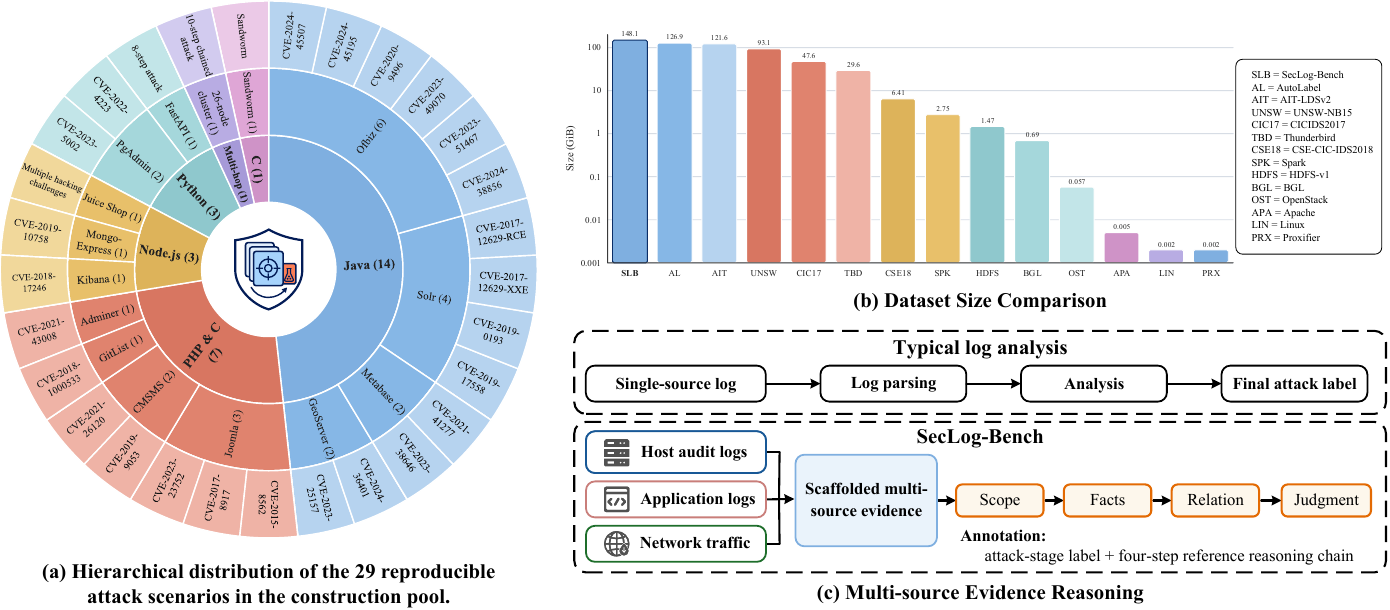}
	\vspace{-1.5ex}
	\caption{Overview of SecLog-Bench data composition and evidence reasoning:
		(a) hierarchy of the 29 reproducible attack scenarios;
		(b) raw-data scale comparison; and
		(c) comparison between typical single-source analysis focused on final attack labels and the SecLog-Bench pipeline for scaffolded multi-source evidence reasoning with attack-stage labels and four-step reference reasoning chains.}
	\label{fig:dataset-overview}
	\vspace{-2ex}
\end{figure*}

This section summarizes the construction pool and final evaluation set of SecLog-Bench. The construction pool contains 29 reproducible attack scenarios, including 25 CVE-based and 4 non-CVE scenarios. The Multi-hop scenario runs in a 26-node service cluster, where the attacker compromises an initial service and then uses compromised services as pivots along an attack path involving ten inter-host hops. Fig.~\ref{fig:dataset-overview}(a) presents the hierarchy of the 29 scenarios by implementation stack, application or scenario family, and individual CVE or non-CVE scenario.

We analyzed the behavior of all 29 source scenarios using MITRE ATT\&CK v17.1~\cite{mitre_attack} and identified 93 associations between scenarios and techniques based on execution evidence, including explicitly observed attempts. These associations cover 23 directly assigned technique and sub-technique IDs and 18 parent technique families, spanning ten ATT\&CK tactics, including Initial Access, Execution, Discovery, Defense Evasion, and Command and Control. Among scenarios with confirmed mappings, we identified 22 distinct sets of directly assigned ATT\&CK IDs and 17 sets after grouping by parent technique. Mean pairwise Jaccard similarity is 0.2904 using direct IDs and 0.3659 at the parent level. Table~\ref{tab:behavior-diversity} presents the direct mappings for ten representative scenarios.

\begin{table}[!t]
	\centering
	\caption{Direct ATT\&CK mappings for ten representative source scenarios.}
	\label{tab:behavior-diversity}
	\footnotesize
	\setlength{\tabcolsep}{5pt}
	\renewcommand{\arraystretch}{1.12}
	\begin{tabularx}{\columnwidth}{@{}p{0.42\columnwidth}>{\raggedright\arraybackslash}X@{}}
		\toprule
		Scenario & Technique and Sub-technique IDs \\
		\midrule
		GeoServer/CVE-2023-25157 & T1190, T1518 \\
		GitList/CVE-2018-1000533 & T1016, T1033, T1059.004, T1083, T1087.001, T1190 \\
		Joomla/CVE-2023-23752 & T1005, T1087, T1190, T1552 \\
		Metabase/CVE-2023-38646 & T1059.007, T1190 \\
		Solr/CVE-2017-12629-XXE & T1087.001, T1190 \\
		Solr/CVE-2019-17558 & T1016, T1033, T1059, T1083, T1087.001, T1190 \\
		OFBiz/CVE-2024-45195 & T1033, T1059.004, T1105, T1190, T1505.003 \\
		Python Demo & T1005, T1059.006, T1078, T1083, T1087.001, T1190, T1552.004 \\
		Multi-hop & T1027.010, T1059, T1059.004, T1090, T1105, T1140, T1190, T1222.002 \\
		Sandworm & T1021.004, T1059.004, T1078, T1105, T1548.003 \\
		\bottomrule
	\end{tabularx}
\end{table}

To examine more diverse multi-hop attack execution patterns, we tested three auxiliary two-target variants: Python Demo $\rightarrow$ GitList, the reverse order, and a forward run with a 90-second inter-target delay. In the forward and reverse runs, the constructed chains preserved the recorded target-entry order, while the connection direction reversed as the entry and pivot roles changed. During the delayed run, normal activity and multi-source log collection continued, and the constructed chain retained 190 service events within the 90-second interval. The subsequent GitList request remained traceable across audit logs, application logs, and network traffic. These variants broaden the range of attack execution patterns tested during construction by varying target order and timing while preserving cross-source evidence traceability.

\begin{table}[!t]
	\centering
	\caption{Scenario-group and stage-label distribution of the final evaluation set. The Scenarios and Windows columns report reproducible construction-pool scenarios and retained evaluation windows, respectively.}
	\label{tab:evaluation-source-distribution}
	\footnotesize
	\setlength{\tabcolsep}{2pt}
	\renewcommand{\arraystretch}{0.95}
	\begin{tabular*}{0.97\columnwidth}{@{\hspace{4pt}\extracolsep{\fill}}ccccccc@{\hspace{4pt}}}
		\toprule
		Scenario group & Scenarios & Windows & Preparation & Execution & Result & SU \\
		\midrule
		Java & 14 & 92 & 20 & 36 & 12 & 24 \\
		PHP \& C & 7 & 58 & 14 & 2 & 27 & 15 \\
		Python & 3 & 29 & 8 & 8 & 9 & 4 \\
		Node.js & 3 & 22 & 9 & 3 & 4 & 6 \\
		Sandworm & 1 & 8 & 0 & 3 & 0 & 5 \\
		Multi-hop & 1 & 7 & 3 & 2 & 2 & 0 \\
		\midrule
		Total & 29 & 216 & 54 & 54 & 54 & 54 \\
		\bottomrule
	\end{tabular*}
\end{table}

The final evaluation set contains 216 filtered, human-reviewed multi-source evidence windows. It is balanced across the four labels, with 54 windows per label. The size of the final evaluation set reflects a deliberate trade-off among annotation depth, manual review effort, label balance, and sample count, rather than a limit on raw-log generation or evidence-window construction. As described in Section~\ref{sec:reference-reasoning-annotation}, every candidate reference chain is manually reviewed against the window evidence and revised as needed, while evidence-source coordinates preserve traceability to the corresponding raw log records. We therefore retain only windows whose labels are supported under the stage criteria of Section~\ref{sec:problem-definition} and the visible evidence boundary. The 54-per-label design provides equal sample support for per-label recall comparisons and complements Macro-F1, which weights the four labels equally by construction. Each retained window carries a four-step human-confirmed reference reasoning chain, yielding 864 annotated reasoning steps in total.

Table~\ref{tab:evaluation-source-distribution} reports the distribution of the final evaluation set by scenario group and stage label. The construction pool includes more Java scenarios, and the Java group therefore contains more retained windows. Window selection is not concentrated on any particular scenario. The label mix within each group reflects the stage evidence exposed by its attack chains.

Table~\ref{tab:dataset-comparison} compares SecLog-Bench with representative public log and network-traffic datasets in terms of source coverage, evidence organization, generation extensibility, reasoning-chain availability, and raw-data scale. Because the table includes both log lines and network-flow records, raw-data scale is reported only as descriptive context. Fig.~\ref{fig:dataset-overview}(b) visualizes the scale comparison.

Across the 29 attack scenarios, each automated construction run completes within 70 minutes, with 20 minutes devoted to raw-log collection and the rest to evidence organization and rule-based preliminary stage labeling. This timing excludes subsequent human review of the evidence windows and their preliminary stage labels. Together with adapter reuse, this per-scenario construction time supports scalability within the evaluated runtime scope. Among the datasets in Table~\ref{tab:dataset-comparison}, SecLog-Bench is the only one that combines multi-source logs, scaffolded evidence organization, extensible scenario generation, and window-level reference reasoning chains.

\begin{table*}[!t]
	\centering
	\caption{Comparison with representative public log and network-traffic datasets. SecLog-Bench provides 216 evidence-window-level reference reasoning chains with 864 annotated reasoning steps.}
	\label{tab:dataset-comparison}
	\footnotesize
	\resizebox{\textwidth}{!}{
		\begin{tabular}{lccccrrrrr}
			\toprule
			Dataset & Multi-source & Evidence organization & Extensibility & Reasoning chains &
			App. log \#lines & Audit log \#lines & Network traffic \#lines & Total \#lines & Size (GiB) \\
			\midrule
			HDFS-v1~\cite{zhu2023loghub} & No & Single-source logs & No & No &
			-- & 11,175,629 & -- & 11,175,629 & 1.470 \\

			Spark~\cite{zhu2023loghub} & No & Single-source logs & No & No &
			-- & 33,236,604 & -- & 33,236,604 & 2.750 \\

			OpenStack~\cite{zhu2023loghub} & No & Single-source logs & No & No &
			-- & 207,820 & -- & 207,820 & 0.057 \\

			BGL~\cite{zhu2023loghub} & No & Single-source logs & No & No &
			-- & 4,747,963 & -- & 4,747,963 & 0.692 \\

			Thunderbird~\cite{zhu2023loghub} & No & Single-source logs & No & No &
			-- & 211,212,192 & -- & 211,212,192 & 29.600 \\

			Linux~\cite{zhu2023loghub} & No & Single-source logs & No & No &
			-- & 25,567 & -- & 25,567 & 0.002 \\

			Apache~\cite{zhu2023loghub} & No & Single-source logs & No & No &
			56,481 & -- & -- & 56,481 & 0.005 \\

			Proxifier~\cite{zhu2023loghub} & No & Single-source logs & No & No &
			21,329 & -- & -- & 21,329 & 0.002 \\

			CICIDS2017~\cite{sharafaldin2018toward} & No & Traffic / flow records & No & No &
			-- & -- & 2,830,743 & 2,830,743 & 47.600 \\

			UNSW-NB15~\cite{moustafa2016unsw} & No & Traffic / flow records & No & No &
			-- & -- & 2,540,044 & 2,540,044 & 93.100 \\

			CSE-CIC-IDS2018 flows~\cite{cse_cic_ids2018} & No & Traffic / flow records & No & No &
			-- & -- & 16,232,943 & 16,232,943 & 6.410 \\

			AIT-LDSv2~\cite{landauer2022aitldsv2} & \textbf{Yes} & Separate multi-source logs & Limited & No &
			N.A. & N.A. & N.A. & N.A. & 121.631 \\

			AutoLabel~\cite{peng2025autolabel} & \textbf{Yes} & Separate multi-source logs & \textbf{Yes} & No &
			N.A. & N.A. & N.A. & N.A. & 126.853 \\

			SecLog-Bench (ours) & \textbf{Yes} & \textbf{Scaffolded evidence} & \textbf{Yes} & \textbf{Yes} &
			166,579 & 157,369,057 & 1,599,414 & 159,135,050 & 148.098 \\
			\bottomrule
		\end{tabular}
	}
	\vspace{0.2em}

	\begin{minipage}{\textwidth}
		\footnotesize
		``Limited'' denotes testbed-based regeneration, where extension to new attack scenarios still requires non-trivial engineering effort.
	\end{minipage}

\end{table*}

Fig.~\ref{fig:dataset-overview}(c) summarizes each benchmark sample: an attack-related evidence window derived from the constructed attack chain and containing scaffolded multi-source evidence from host audit logs, application logs, and network traffic, an attack-stage label verified against auditable criteria, and a four-step reference reasoning chain. Together, the structured evidence and reasoning-oriented annotation make SecLog-Bench a reasoning-ready benchmark for controlled evaluation of window-level attack-stage prediction and reasoning reliability, rather than for model training or large-scale log-detection evaluation.

\section{LLM Performance}
\label{sec:experiments}

Following the protocol in Section~\ref{sec:evaluation-methodology}, this section benchmarks LLM performance on attack-stage reasoning over scaffolded multi-source evidence along four dimensions: attack-stage prediction, reasoning quality and prediction--reasoning consistency, failure modes, and the effects of evidence organization and source coverage. We also examine sensitivity to sampling and voting settings.

\subsection{Experimental Setup}
\label{sec:experimental-setup}
We evaluate eight representative LLMs: Qwen3-8B~\cite{qwen_technical_report}, Qwen3-32B~\cite{qwen_technical_report}, Llama-3.1-8B-Instruct~\cite{meta2024llama3}, DeepSeek-R1-0528~\cite{deepseek2025r1}, Qwen3.6-Plus~\cite{qwen36plus_official}, Gemini 3.1 Pro Preview~\cite{google_gemini31pro}, Claude Sonnet 4.6~\cite{anthropic_claude_sonnet46}, and GPT-5.4~\cite{openai_gpt54}. The selected models span different model families and capability levels, enabling comparison of their performance on multi-source log attack-stage reasoning. The closed-source models are accessed through their official service interfaces. Among the open-source models, DeepSeek-R1-0528 is accessed via a remote service interface, whereas the others are deployed on a local server. The local server runs Ubuntu 22.04.5 LTS with two 32-GB NVIDIA RTX 5090 GPUs, 25 Intel Xeon Platinum 8470Q vCPUs, and 120 GB of system memory.

To improve comparability, all models are evaluated using the same samples, evidence view, output requirements, and metrics. The main evaluation runs were conducted between May 21 and June 27, 2026, with temperature $=0$, top-$p=1$, and model-specific maximum output lengths ranging from 2,048 to 8,192 tokens. For the locally deployed models, decoding top-$k$ truncation was disabled (top-$k=0$). No explicit reasoning-effort parameter was supplied, so provider-default reasoning behavior was used where applicable. Requests that encountered transient failures were retried using the same inference configuration as in the initial attempt. The reported cross-model differences therefore reflect performance under each model's available inference interface rather than a comparison with matched reasoning budgets. The resulting metrics characterize model performance only on this fixed set of window-level samples and do not demonstrate generalization to unseen attack scenarios. The benchmark data and experimental code are publicly available at \url{https://github.com/SecLog-Open/SecLog}.

\subsection{Attack-Stage Prediction Performance}
\label{sec:rq1-stage-prediction}

We compare attack-stage prediction performance across models using the same scaffolded, compressed multi-source evidence view (Multi-Comp-Scaf.). Table~\ref{tab:stage-prediction} reports Accuracy, Macro-F1, and per-label recall.

\begin{table*}[!t]
	\centering

	\caption{Attack-stage prediction performance and per-label recall. Prep., Exec., Res., and SU denote the preparation, execution, result, and SU labels, respectively. Best and second-best results are highlighted in dark and light green, respectively.}
	\label{tab:stage-prediction}
	\footnotesize
	\setlength{\tabcolsep}{2.0mm}
	\begin{tabular}{lcccccc}
		\toprule
		Model & Accuracy$\uparrow$ & Macro-F1$\uparrow$ & Prep. Recall$\uparrow$ & Exec. Recall$\uparrow$ & Res. Recall$\uparrow$ & SU Recall$\uparrow$ \\
		\midrule
		Qwen3-8B & 23.15\% & 19.27\% & 3.70\% & 38.89\% & 7.41\% & 42.59\% \\
		Qwen3-32B & 26.39\% & 25.95\% & 12.96\% & \best{50.00\%} & 24.07\% & 18.52\% \\
		Llama-3.1-8B-Instruct & 24.07\% & 20.03\% & \best{62.96\%} & 16.67\% & 9.26\% & 7.41\% \\
		DeepSeek-R1 & 27.31\% & 26.77\% & 12.96\% & 14.81\% & \best{46.30\%} & 35.19\% \\
		Qwen3.6-Plus & 33.80\% & 34.98\% & 37.04\% & 29.63\% & 31.48\% & 37.04\% \\
		Gemini 3.1 Pro Preview & \second{37.50\%} & \second{37.17\%} & 31.48\% & 44.44\% & 24.07\% & \best{50.00\%} \\
		Claude Sonnet 4.6 & \best{44.44\%} & \best{45.07\%} & \second{50.00\%} & \second{48.15\%} & \second{33.33\%} & 46.30\% \\
		GPT-5.4 & 33.33\% & 33.29\% & 22.22\% & 35.19\% & 27.78\% & \second{48.15\%} \\
		\midrule
		Mean & -- & -- & 29.17\% & 34.72\% & 25.46\% & 35.65\% \\
		\bottomrule
	\end{tabular}
\end{table*}

Table~\ref{tab:stage-prediction} shows that current LLMs achieve limited attack-stage prediction performance under the compressed multi-source evidence view. Claude Sonnet 4.6 obtains the best overall result, with 44.44\% Accuracy and 45.07\% Macro-F1. This result exceeds the 25\% uniform-random reference for four labels, but fewer than half of the evidence windows receive the correct stage label. Thus, the task remains difficult even after the model is given attack-related evidence and does not need to detect attacks from open-ended logs.

The per-label recall results further show that model behavior is uneven across stage labels. No model consistently distinguishes all four labels: high recall on one label does not translate into balanced performance across the others. For example, Llama-3.1-8B-Instruct has the highest Prep. Recall but only 20.03\% Macro-F1, while DeepSeek-R1 has the highest Res. Recall but low Exec. Recall; across models, \stage{result} has the lowest mean recall at 25.46\%, a weakness examined further in Section~\ref{sec:rq3-failure-modes}.

In the parameter-scale comparison between Qwen3-8B and Qwen3-32B, the 32B model achieves higher Accuracy and Macro-F1 but lower SU recall, indicating that its overall prediction advantage is not uniform across labels.

\noindent\textbf{Finding 1:} Current LLMs can identify some attack-stage signals from scaffolded multi-source evidence, but their attack-stage prediction performance remains limited. The main difficulty is stable discrimination among concrete stage labels and the evidence-insufficient SU label, rather than detecting whether an attack exists.

\subsection{Reasoning Quality and Prediction--Reasoning Consistency}
\label{sec:rq2-reasoning-quality}

\begin{table*}[!t]
	\centering

	\caption{Reasoning quality, prediction--reasoning consistency, and efficiency across models. Avg. Tok. denotes average reasoning length, and SRF1/1K Tok. denotes SRF1 per 1K output tokens. $\diamond$ denotes a diagnostic metric with no inherently preferred direction.}
	\label{tab:reasoning-quality}
	\footnotesize
	\setlength{\tabcolsep}{3.2pt}

	\begin{tabular}{lccccccccccc}
		\toprule
		\multirow{2}{*}{Model}
		& \multicolumn{3}{c}{Reasoning Quality}
		& \multicolumn{6}{c}{Prediction--Reasoning Consistency}
		& \multicolumn{2}{c}{Efficiency} \\
		\cmidrule(lr){2-4}
		\cmidrule(lr){5-10}
		\cmidrule(lr){11-12}
		& SRP$\uparrow$
		& SCC$\uparrow$
		& SRF1$\uparrow$
		& TT$\uparrow$
		& TF$\downarrow$
		& FT$^{\diamond}$
		& FF$\downarrow$
		& TT/Correct$\uparrow$
		& FT/Wrong$^{\diamond}$
		& SRF1/1K Tok.$\uparrow$
		& Avg. Tok. \\
		\midrule

		Qwen3-8B
		& 32.27\% & 56.05\% & 40.96\%
		& 12.50\% & \second{10.65\%} & 19.44\% & 57.41\%
		& 54.00\% & 25.30\%
		& 84.99 & 481.92 \\

		Qwen3-32B
		& 29.26\% & 52.14\% & 37.49\%
		& 12.50\% & 13.89\% & 18.06\% & 55.56\%
		& 47.37\% & 24.53\%
		& 109.33 & 342.92 \\

		Llama-3.1-8B-Instruct
		& 32.65\% & 58.58\% & 41.93\%
		& 9.26\% & 14.81\% & 25.00\% & 50.93\%
		& 38.46\% & 32.93\%
		& \second{136.11} & 308.07 \\

		DeepSeek-R1
		& 26.73\% & 51.55\% & 35.21\%
		& 15.74\% & 11.57\% & 13.43\% & 59.26\%
		& 57.63\% & 18.47\%
		& 83.15 & 423.43 \\

		Qwen3.6-Plus
		& 28.84\% & 65.10\% & 39.97\%
		& 17.13\% & 16.67\% & 16.20\% & 50.00\%
		& 50.68\% & 24.48\%
		& 88.09 & 453.76 \\

		Gemini 3.1 Pro Preview
		& \second{33.53\%} & 62.05\% & \second{43.54\%}
		& \second{24.54\%} & 12.96\% & 14.81\% & \second{47.69\%}
		& \second{65.43\%} & 23.70\%
		& 99.30 & 438.45 \\

		Claude Sonnet 4.6
		& 25.93\% & \second{76.08\%} & 38.67\%
		& 18.52\% & 25.93\% & 5.56\% & 50.00\%
		& 41.67\% & 10.00\%
		& 41.41 & 933.89 \\

		GPT-5.4
		& \best{42.61\%} & \best{77.04\%} & \best{54.87\%}
		& \best{27.31\%} & \best{6.02\%} & 34.72\% & \best{31.94\%}
		& \best{81.94\%} & 52.08\%
		& \best{144.25} & 380.39 \\

		\bottomrule
	\end{tabular}
	\vspace{-2ex}
\end{table*}

\begin{figure}[!t]
	\centering
	\includegraphics[width=0.86\columnwidth]{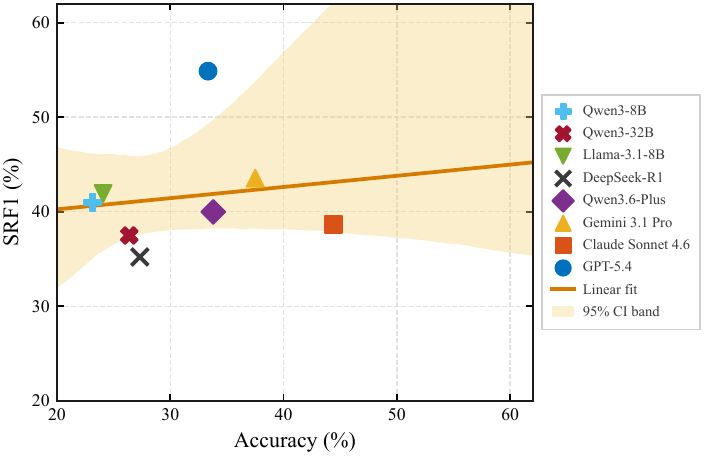}
	\vspace{-1.5ex}
	\caption{Model-level relationship between attack-stage prediction accuracy and reasoning quality. Each point denotes one model. The orange line shows the least-squares fit, and the yellow band denotes the 95\% bootstrap confidence interval of the fitted trend. The Pearson correlation is weak ($r=0.15$, 95\% CI $[-0.62, 0.77]$).}
	\label{fig:accuracy-srf1}
	\vspace{-1ex}
\end{figure}

In this subsection, we examine whether correct attack-stage predictions are accompanied by reliable reasoning. Table~\ref{tab:reasoning-quality} summarizes reasoning quality, prediction--reasoning consistency, and efficiency across models, while Fig.~\ref{fig:accuracy-srf1} analyzes the model-level relationship between Accuracy and SRF1.

Table~\ref{tab:reasoning-quality} shows that attack-stage prediction correctness is not equivalent to reasoning reliability. If the two were aligned, the best-performing model in attack-stage prediction would also be consistently strong in SRF1 and TT/Correct. Claude Sonnet 4.6 obtains the best attack-stage prediction performance, as reported in Section~\ref{sec:rq1-stage-prediction}, but its SRF1 is 38.67\% and its TT/Correct is 41.67\%, indicating that many of its correct predictions do not satisfy the reasoning reliability criteria. For example, it correctly predicts \stage{execution} for a window containing a template-injection request followed by process forks, but incorrectly treats the forks as direct evidence that injected commands were executed, although no such commands are visible. In contrast, GPT-5.4 does not have the highest Accuracy, but it has the highest SRF1, 54.87\%, the highest TT/Correct, 81.94\%, and the lowest TF, 6.02\%. The mismatch is bidirectional: a model may reach a correct label with unreliable reasoning, while GPT-5.4's FT/Wrong rate of 52.08\% shows that incorrect labels can also accompany reasoning that meets the reliability criteria. Across all eight models, SCC exceeds SRP; for example, Claude Sonnet 4.6 obtains 76.08\% SCC but only 25.93\% SRP, suggesting broader stage-condition coverage than claim-level precision, although part of the gap may reflect the different metric constructions. Attack-stage prediction accuracy alone therefore provides an incomplete measure of a model's reasoning reliability.

To examine whether the proposed reasoning metrics distinguish reliable outputs, we compare SRP, SCC, and SRF1 with ROUGE-L F1~\cite{lin2004rouge} and BERTScore F1~\cite{zhang2020bertscore} on 352 human-assessed outputs from 44 windows (Table~\ref{tab:reference-metric-validation}). For both reference-similarity baselines, each model output is compared against its corresponding reference reasoning chain. SRF1 achieves the highest area under the receiver operating characteristic curve (AUROC) of 0.918 and average precision (AP) of 0.933, compared with approximately 0.62 AUROC and 0.612 AP for the reference-similarity baselines. This comparison assesses alignment with human judgments under the reliability rubric in Section~\ref{sec:evaluation-methodology}. BERTScore's 512-token input limit truncated seven candidate outputs and 336 paired reference inputs, limiting its suitability for measuring similarity to long reference reasoning chains. In contrast, ROUGE-L was computed on the full texts.

\begin{table}[!t]
	\centering
	\caption{Comparison of the proposed reasoning metrics with conventional reference-similarity metrics.}
	\label{tab:reference-metric-validation}
	\footnotesize
	\setlength{\tabcolsep}{2pt}
	\renewcommand{\arraystretch}{0.95}
	\begin{tabular*}{0.97\columnwidth}{@{\hspace{4pt}\extracolsep{\fill}}lcc@{\hspace{4pt}}}
		\toprule
		Metric & AUROC (95\% CI) & AP (95\% CI) \\
		\midrule
		ROUGE-L F1 & 0.620 [0.557, 0.681] & 0.612 [0.521, 0.708] \\
		BERTScore F1 & 0.618 [0.542, 0.688] & 0.612 [0.524, 0.700] \\
		SRP & 0.901 [0.869, 0.930] & 0.908 [0.868, 0.942] \\
		SCC & 0.853 [0.813, 0.893] & 0.851 [0.789, 0.902] \\
		SRF1 & \textbf{0.918 [0.886, 0.947]} & \textbf{0.933 [0.904, 0.958]} \\
		\bottomrule
	\end{tabular*}
	\par\vspace{0.8ex}
	\noindent\parbox{\columnwidth}{\scriptsize
		The 95\% confidence intervals (CIs) are based on 10,000 stage-stratified bootstrap resamples of windows, retaining all eight model outputs per sampled window.
	}
	\vspace{-2ex}
\end{table}

For efficiency, Avg. Tok. measures average reasoning length, while SRF1/1K Tok. measures SRF1 per 1K output tokens. Claude Sonnet 4.6 produces the longest reasoning outputs, with 933.89 tokens on average, but has the lowest SRF1/1K Tok., 41.41; GPT-5.4 produces shorter reasoning outputs, with 380.39 tokens on average, and reaches the highest SRF1/1K Tok., 144.25. This suggests that longer reasoning outputs alone do not guarantee more reliable reasoning.
In the parameter-scale comparison between Qwen3-8B and Qwen3-32B, the 32B model is more token-efficient but does not achieve higher reasoning reliability.

Fig.~\ref{fig:accuracy-srf1} visualizes this relation across all models. The model points do not show a clear positive relationship between Accuracy and SRF1: GPT-5.4 attains the highest SRF1 despite a mid-range Accuracy, whereas Claude Sonnet 4.6 has the highest Accuracy but an SRF1 of only 38.67\%. Consistent with this scatter, the Pearson correlation across the eight evaluated models is weak ($r=0.15$, 95\% CI $[-0.62, 0.77]$, $n=8$) and does not indicate a clear association between Accuracy and reasoning quality.

We also reapply the reliability decision rule to the 1{,}728 sample-level outputs at SRP thresholds of 0.30 (Loose), 0.33 (Main), and 0.36 (Strict), with the SCC and SRF1 thresholds and the two hard gates fixed. The reliability rate (37.1\%, 35.6\%, 31.1\%) decreases as the SRP threshold becomes stricter, whereas the overall mismatch rate (32.8\%, 32.5\%, 31.4\%) remains relatively stable, indicating that the prediction--reasoning mismatch is not sensitive to the choice of SRP threshold within this range.

\noindent\textbf{Finding 2:}
Correct attack-stage predictions may still be accompanied by unreliable reasoning. Therefore, final-label accuracy should be complemented by measures of reasoning quality and prediction--reasoning consistency.

\subsection{Failure Modes in Attack-Stage Reasoning}
\label{sec:rq3-failure-modes}

\begin{figure*}[!t]
	\centering
	\includegraphics[width=\textwidth]{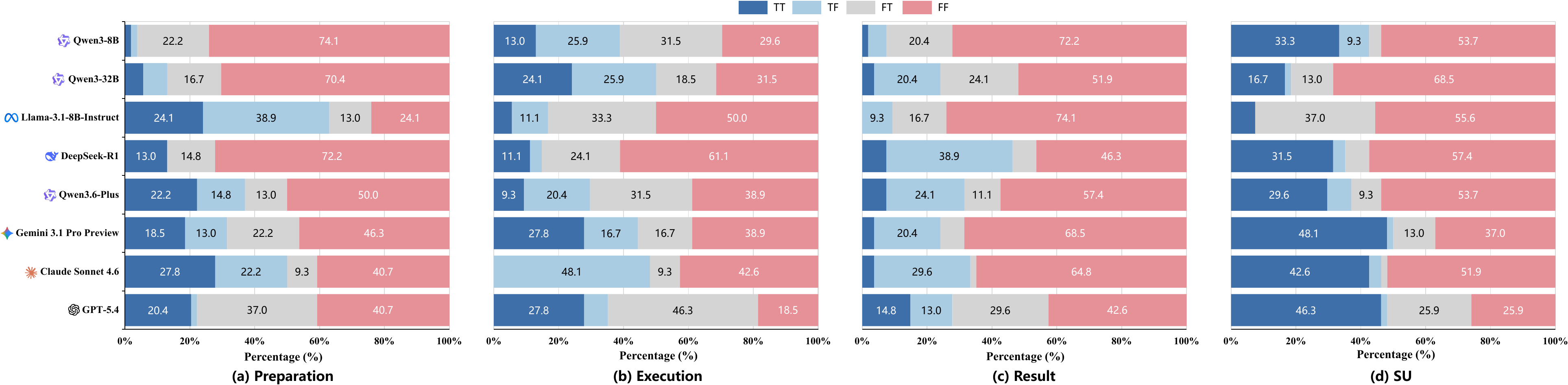}
	\vspace{-3ex}
	\caption{Label-wise prediction--reasoning consistency. Each subfigure groups TT/TF/FT/FF outcomes by the ground-truth stage label $y$, following the definitions in Section~\ref{sec:evaluation-methodology}.}
	\label{fig:label-wise-consistency}
	\vspace{-2ex}
\end{figure*}

\begin{figure}[!t]
	\centering
	\includegraphics[width=0.45\textwidth]{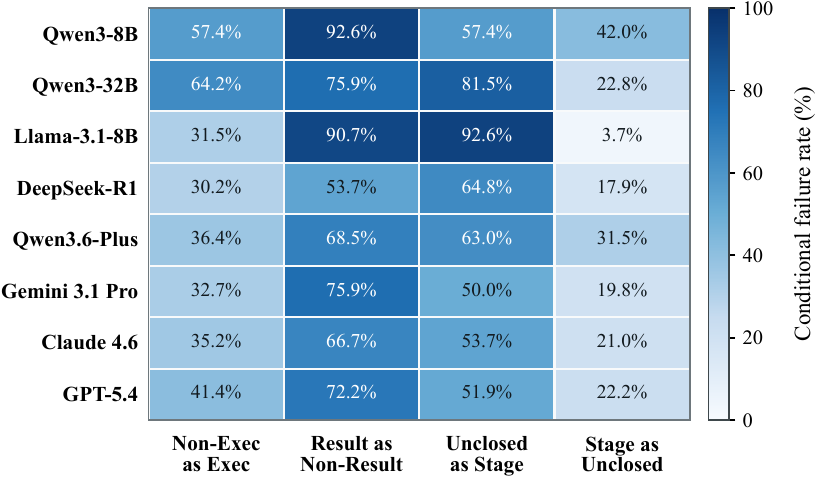}
	\caption{Conditional failure patterns across models. The columns report failure rates for non-\stage{execution} windows predicted as \stage{execution} (Non-Exec as Exec), \stage{result} windows predicted as non-\stage{result} labels (Result as Non-Result), SU windows predicted as concrete stages (Unclosed as Stage), and concrete-stage windows predicted as SU (Stage as Unclosed).}
	\label{fig:failure-modes}
\end{figure}

We analyze where attack-stage reasoning failures concentrate. Fig.~\ref{fig:label-wise-consistency} examines prediction--reasoning consistency by ground-truth stage label, while Fig.~\ref{fig:failure-modes} summarizes four conditional failure patterns. We focus on whether models correctly determine whether the visible evidence supports a concrete attack-stage judgment or instead warrants the SU label.

Fig.~\ref{fig:label-wise-consistency} shows that per-label recall can conceal whether correct predictions are backed by reliable reasoning. For \stage{preparation}, Llama-3.1-8B-Instruct's 62.96\% recall splits into only 24.1\% TT and 38.9\% TF. The \stage{result} label is the weakest case: TT remains no higher than 14.8\% for all models, while FF exceeds 50\% for most models. This indicates that models often miss visible consequence evidence. For SU, FF exceeds 50\% for six of the eight models, so evidence-insufficient windows are not handled consistently. The starkest case is \stage{execution}: Claude Sonnet 4.6 reaches 48.15\% recall, yet Fig.~\ref{fig:label-wise-consistency}(b) shows 0\% TT and 48.1\% TF, so no correct \stage{execution} prediction is backed by reliable reasoning.

Fig.~\ref{fig:failure-modes} shows that failures are concentrated in a few recurring patterns. The most persistent pattern is missing \stage{result} evidence: Result as Non-Result, which equals one minus the \stage{result} recall reported in Table~\ref{tab:stage-prediction}, is high for every model, ranging from 53.7\% to 92.6\%. This means that models frequently fail to recognize when visible consequences, such as responses, returned data, or state changes, already support the \stage{result} label. The second major pattern is predicting concrete stage labels for evidence-insufficient windows. Unclosed as Stage is at least 50.0\% for all models and reaches 92.6\% for Llama-3.1-8B-Instruct, showing a broad tendency to over-commit when the evidence should remain SU.

Other failures are more model-specific. Qwen3-32B has the highest Non-Exec as Exec rate, 64.2\%, indicating a tendency to prematurely interpret non-\stage{execution} evidence as \stage{execution}. For example, it predicts \stage{execution} for an SU window with repeated SSH \texttt{NET\_ACCEPT} and \texttt{PROCESS\_FORK} events but no visible command execution, file write, or state change, conflating service handling with attack execution. SU calibration fails in opposite directions: Qwen3-8B has the highest Stage as Unclosed rate, mapping 42.0\% of concrete-stage windows to SU, whereas Llama-3.1-8B-Instruct shows the reverse pattern: despite the high Unclosed as Stage rate noted above, it maps only 3.7\% of concrete-stage windows to SU. Overall, these patterns suggest that final-label errors often reflect a deeper difficulty in determining whether the visible evidence is sufficient to support a concrete stage judgment.

\noindent\textbf{Finding 3:}
	The main failures concern whether visible evidence supports a concrete attack-stage judgment or warrants an SU label. Models frequently miss visible \stage{result} consequences and often assign concrete stage labels to evidence-insufficient windows. Some models also prematurely assign \stage{execution} or overuse SU, showing that current LLMs still struggle to calibrate evidence sufficiency for attack-stage reasoning.

\subsection{Effects of Evidence Organization and Source Coverage}

\label{sec:rq4-ablation}

\begin{figure}[!t]
	\centering

	\subfloat[Evidence-view ablation\label{fig:evidence-view-ablation}]{%
		\includegraphics[width=0.49\columnwidth]{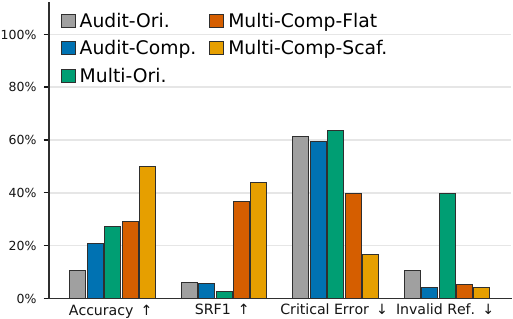}%
	}\hfill
	\subfloat[Source-combination ablation\label{fig:source-combination-ablation}]{%
		\includegraphics[width=0.49\columnwidth]{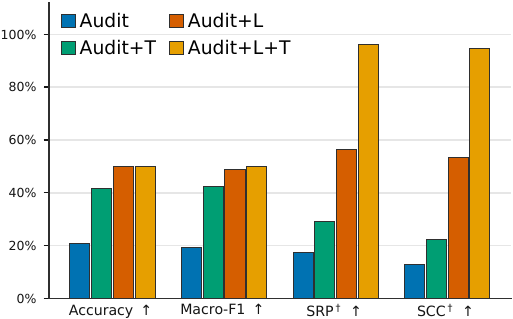}%
	}

	\caption{Ablation results for evidence organization and source coverage.
		(a) compares audit-only and multi-source settings under original and compressed evidence views, with flat and scaffolded variants for compressed multi-source evidence.
		(b) compares source combinations under the scaffolded compressed view.
		Invalid Ref. denotes Invalid Evidence Reference; Ori. and Comp. denote uncompressed and compressed evidence views, respectively; Flat and Scaf. denote flat and scaffolded presentations; L and T denote application logs and network traffic.
		SRP$^\dagger$ and SCC$^\dagger$ are source-availability diagnostics computed against the full-view reference checklist.}
	\label{fig:evidence-organization-ablation}
	\vspace{-3ex}
\end{figure}

In this subsection, we evaluate how the organization and source coverage of model-visible evidence influence attack-stage reasoning. We fix the evaluated model to GPT-5.4 and compare different evidence settings on the same subset of 96 evidence windows, so that the results primarily reflect evidence organization and source coverage rather than model or sample differences. Fig.~\ref{fig:evidence-organization-ablation} presents two controlled ablations: (a) examines the effects of evidence compression and presentation across audit-only and multi-source settings, while (b) evaluates source coverage under the scaffolded compressed view.

Fig.~\ref{fig:evidence-organization-ablation}(a) shows higher stage-prediction performance for the compressed audit-only view and the scaffolded compressed multi-source view than for their corresponding original views. Accuracy rises from 10.42\% to 20.83\% for audit-only evidence and from 27.08\% to 50.00\% for multi-source evidence with scaffolded presentation. Beyond prediction accuracy, Multi-Ori. achieves only 2.77\% SRF1, whereas Multi-Comp-Scaf. reaches the highest SRF1 (44.09\%) and also reduces the Critical Error rate. This contrast suggests that large volumes of poorly organized multi-source evidence can hinder reliable reasoning, as redundant events, weakly related fragments, and heterogeneous formats make stage-relevant evidence harder to identify and connect. The compressed, scaffolded view helps mitigate these difficulties by organizing evidence according to the salience, cross-source continuity, and coverage principles described in Section~\ref{sec:reference-reasoning-annotation}.

To examine the role of presentation, we use Multi-Comp-Flat, which presents the same compressed multi-source evidence as Multi-Comp-Scaf. in a flat sequence rather than in a scaffolded form. Compared with Multi-Ori., Multi-Comp-Flat reduces the invalid-reference rate but yields only a modest Accuracy gain. Scaffolded presentation further raises Accuracy from 29.17\% to 50.00\% and SRF1 from 36.59\% to 44.09\%, while reducing the Critical Error rate from 39.58\% to 16.67\%. Invalid-reference rates remain similar (5.21\% and 4.17\%). These gains highlight the effect of evidence presentation on measured performance, with the model and underlying evidence held fixed.

Fig.~\ref{fig:evidence-organization-ablation}(b) further shows that source coverage and evidence organization are complementary. Under the scaffolded compressed view, audit logs alone provide a host-side behavioral skeleton, but remain limited for attack-stage reasoning. Adding traffic supplies endpoint, direction, and transfer context, while adding application logs improves source-availability diagnostics by exposing request, handler, and response semantics that are difficult to infer from audit evidence alone. Combining all three sources achieves the best Macro-F1 and the strongest source-availability diagnostics, with 96.33\% SRP$^\dagger$ and 94.78\% SCC$^\dagger$. These results show that correlating host, application, and network evidence provides the model with more complete context for distinguishing isolated suspicious events from coherent attack progress.

\noindent\textbf{Finding 4:}
Evidence compression, scaffolded presentation, and source coverage make \emph{multi-source evidence} more usable for attack-stage reasoning. Compression retains stage-relevant evidence with traceable coordinates, scaffolded presentation improves stage judgments and reduces critical errors with the underlying evidence held fixed, and broader source coverage supplies correlated host, application, and network context. Overall, Multi-Comp-Scaf. achieves the strongest attack-stage prediction performance among the evaluated evidence views.

\subsection{Sensitivity to Sampling and Voting Settings}
\label{sec:sampling-sensitivity}
We assess sensitivity to sampling and voting settings for Qwen3-8B and Llama-3.1-8B-Instruct using newly generated answers on the main evaluation set (Table~\ref{tab:sampling-sensitivity}). All runs use top-$p=1$ and an 8,192-token output cap. Temperature 0 serves as the single-generation baseline. At nonzero temperatures, we use seeds 1001--1005 to generate five answers and apply plurality voting to the first $n\in\{1,3,5\}$ answers. The most frequent valid label is selected, with ties broken by seed order. After retries, all evaluation windows are scored using the final answers, and windows without a valid vote are counted as incorrect.

\begin{table*}[!t]
	\centering
	\caption{Effects of sampling and voting settings on attack-stage prediction. Entries are Accuracy / Macro-F1 (\%). Here, $n$ denotes the number of answers used for voting. Off denotes decoding top-$k=0$ (disabled), and --- denotes not evaluated.}
	\label{tab:sampling-sensitivity}
	\footnotesize
	\setlength{\tabcolsep}{5pt}
	\begin{tabular*}{\textwidth}{@{\extracolsep{\fill}}lccccccc@{}}
		\toprule
		\multirow{2}{*}{Temperature} & \multirow{2}{*}{\shortstack{Decoding\\top-$k$}} & \multicolumn{3}{c}{Qwen3-8B} & \multicolumn{3}{c}{Llama-3.1-8B-Instruct} \\
		\cmidrule(lr){3-5}\cmidrule(lr){6-8}
		& & $n=1$ & $n=3$ & $n=5$ & $n=1$ & $n=3$ & $n=5$ \\
		\midrule
		0.0 & Off & 23.15 / 19.27 & --- & --- & 24.07 / 20.03 & --- & --- \\
		0.3 & Off & 18.06 / 15.53 & 22.22 / 18.86 & 21.76 / 18.69 & 21.30 / 19.92 & 18.52 / 17.11 & 18.06 / 16.16 \\
		0.7 & Off & 22.69 / 19.36 & 21.30 / 17.66 & 24.54 / 19.87 & 21.30 / 20.92 & 21.76 / 20.25 & 20.37 / 18.54 \\
		0.7 & 20 & 23.61 / 20.00 & 22.69 / 18.58 & 25.00 / 20.50 & 21.30 / 21.32 & 21.76 / 20.29 & 20.37 / 18.56 \\
		\bottomrule
	\end{tabular*}
\end{table*}

Across the tested settings, multi-generation voting does not consistently improve prediction: increasing $n$ from 1 to 3 at temperature 0.3 raises Accuracy for Qwen but lowers it for Llama. With decoding top-$k$ disabled, raising temperature from 0.3 to 0.7 improves Accuracy at $n=5$ for both models. At temperature 0.7, decoding top-$k=20$ yields at most 1.39 percentage points of Accuracy gain for Qwen and none for Llama. The highest Accuracy is 25.00\% for Qwen and 24.07\% for Llama, while the highest Macro-F1 is 20.50\% and 21.32\%, respectively. Compared with the zero-temperature baseline, the maximum Accuracy gain is only 1.85 percentage points for Qwen, with no gain for Llama. Overall, the tested sampling and voting settings do not provide a consistent improvement in stage-prediction performance.

\section{Discussion and Limitations}
\label{sec:discussion-limitations}
\subsection{Discussion}

Our results show that inferring attack progress from cybersecurity logs is not merely a problem of classifying attack stages, but a security reasoning task. For reliable use in this task, a model should make a stage judgment supported by the available evidence, ground its reasoning in visible log evidence, and avoid assigning a concrete stage when the evidence is insufficient. Although current LLMs do not yet meet these requirements reliably, they may assist analysts in organizing facts from multiple log sources, identifying possible cues about attack progress, and exposing evidence gaps. Their predictions and reasoning outputs still require human review.

The experiments also show that structured multi-source evidence views make log evidence more usable for reasoning about attack progress. Under the Multi-Ori. view, presenting uncompressed multi-source evidence exposes the model to weakly relevant fragments, making it harder to locate key evidence and generate valid evidence references. Therefore, LLM input design for cybersecurity log reasoning should prioritize evidence traceability and clear cross-source relations, rather than simply expanding context length. Future work could examine how evidence retrieval and organization can be combined to support reliable reasoning over larger, less curated log streams.

\subsection{Limitations}

SecLog-Bench evaluates this task using attack-related evidence windows selected by its construction pipeline. Within these windows, \stage{stage\_unclosed} denotes insufficient evidence for a concrete stage judgment rather than benign activity. The results therefore characterize post-detection attack-progress reasoning and do not evaluate attack discovery and candidate-window retrieval from open log streams, benign-versus-malicious discrimination, or operational false-positive rates for attack detection.

The study also has the following limitations:

\begin{enumerate}
	\item \textbf{Evaluation scale and scenario coverage.}
	The main evaluation set contains 216 human-reviewed multi-source evidence windows, reflecting a deliberate trade-off among annotation depth, label balance, and sample count. This design supports controlled analysis of attack-stage reasoning, but the current evaluation set cannot fully capture the complexity of enterprise environments. Its balanced stage-label distribution also does not represent operational class prevalence. These limitations concern the scale and composition of the current reasoning-annotated evaluation set rather than the extensibility of the log-generation and evidence-window construction pipeline. The auxiliary indices and cross-source alignment process can be adapted to new reproducible attack scenarios.

	\item \textbf{AI-assisted annotation and assessment.}
	For reference reasoning, AI is used only to generate candidate chains and flag review risks; all reference chains are manually reviewed. For reasoning-quality assessment, evaluation assistants are used only for intermediate checks; final reliability decisions are made by fixed local scoring rules. Human-only validation shows substantial agreement with the reliability decisions of the fixed evaluation pipeline. Together, these safeguards limit the impact of possible AI-assistant errors or biases. However, exposure to AI-generated candidate chains and risk flags may still introduce anchoring bias during human review.

	\item \textbf{Stage-label granularity.}
	We use four labels to characterize window-level attack progress. This label space supports controlled comparison, but abstracts away finer-grained transitions in real attack lifecycles.

	\item \textbf{Model versions and reproducibility.}
	We evaluate service-based and locally deployed models. For service-based models, reproducibility may be affected by model updates, API changes, and varying inference settings.

\end{enumerate}

\section{Related Work}
\label{sec:related-work}

\subsection{Attack Evidence Organization}

Security-log analysis methods represent and organize log evidence at different levels of abstraction. Conventional log-analysis pipelines commonly derive templates through log parsing~\cite{he2017drain,khan2024impact,jiang2024loghub2} and model template or event sequences for anomaly detection~\cite{du2017deeplog,chai2025logsequence}. Provenance-based systems~\cite{hossain2017sleuth,milajerdi2019holmes,hassan2019nodoze,hassan2020omegalog,li2021threat,xu2022depcomm,inam2023sok} instead construct and organize attack evidence from audit logs using techniques such as dependency modeling, provenance triage, and graph summarization. Within this provenance-oriented line, prior work has explored provenance tracing~\cite{ma2016protracer}, runtime detection of advanced persistent threats~\cite{han2020unicorn}, automated attack investigation~\cite{atlas2021attackinvestigation,ding2023airtag}, and context-aware alert prioritization~\cite{liu2022rapid}.

Complementary approaches broaden attack evidence by analyzing network communication patterns, interpreting log semantics, and aligning heterogeneous log sources. BotMiner~\cite{gu2008botminer}, for example, correlates network communication and activity patterns to detect botnets without relying on a specific command-and-control protocol or botnet structure. WATSON~\cite{zeng2021watson} introduces semantic interpretation for audit logs, unified semantic parsing~\cite{tan2024unified} supports semantic normalization of heterogeneous logs, and AutoLabel~\cite{peng2025autolabel} aligns application logs, traffic logs, and audit logs for fine-grained attack labeling. Taken together, these approaches support attack reconstruction, threat detection, incident investigation, dependency tracking, and attack labeling, but do not specifically address the organization of correlated evidence into LLM-facing views for inferring attack progress from logs.

\subsection{LLM Reasoning}

LLM-based log-analysis studies~\cite{liu2024logprompt,liu2025loglm,zhang2025llmlade,ma2026llm4log} have explored instruction-based log understanding, anomaly detection with explanation, and broader LLM applications in log analysis. Reasoning-quality benchmarks such as MME-CoT~\cite{jiang2025mmecot} and M3CoTBench~\cite{jiang2026m3cotbench} show that final-answer correctness can hide flawed intermediate reasoning~\cite{golovneva2023roscoe}, while security-oriented LLM benchmarks~\cite{deason2025cybersoceval,ctibench2024,cherif2025dfirmetric,ma2026attackseqbench} evaluate models on security operations, cyber threat intelligence, digital forensics, and attack-behavior analysis. These two directions rarely meet on cybersecurity log evidence: reasoning benchmarks are not designed for security-log analysis, while security benchmarks usually emphasize task performance rather than reasoning.

\section{Conclusion}
\label{sec:conclusion}

We propose SecLog-Bench, a benchmark for evaluating cybersecurity log reasoning in post-detection threat analysis. SecLog-Bench constructs scaffolded multi-source evidence from host audit logs, application logs, and network traffic, and evaluates whether LLMs can infer attack progress, predict attack-stage labels, and produce reasoning consistent with reference reasoning chains.

Experiments on eight representative LLMs over 216 scaffolded multi-source evidence windows show that current models remain limited in inferring attack progress from log evidence: attack-stage prediction is difficult, and correct predictions do not necessarily imply reliable reasoning. Ablation results show that scaffolded multi-source evidence is most useful when raw logs are compressed into stage-relevant evidence and supported by complementary host, application, and network context. SecLog-Bench provides a systematic basis for evaluating cybersecurity log reasoning beyond final-label accuracy.

\bibliographystyle{IEEEtran}
\bibliography{references}

\balance

\end{document}